\documentclass[usenatbib,usegraphicx]{mn2e}

\usepackage{natbib}
\usepackage{amsmath}
\usepackage[]{hyperref}
\usepackage{graphicx}
\usepackage[dvipsnames]{xcolor}
\usepackage{units}
\usepackage{ulem}
\usepackage{fixltx2e}
\DeclareGraphicsExtensions{.eps,.pdf,.png,.jpg}
\hypersetup{
    colorlinks=true,
    linkcolor=black,
    citecolor=black,
    filecolor=black,
    urlcolor=black,
}

\usepackage{deluxetable}
\usepackage{color}

\def\lesssim{\mathrel{\hbox{\rlap{\hbox{\lower3pt\hbox{$\sim$}}}\hbox{\raise2pt\hbox{$<$}}}}}
\def\gtrsim{\mathrel{\hbox{\rlap{\hbox{\lower3pt\hbox{$\sim$}}}\hbox{\raise2pt\hbox{$>$}}}}}

\def\chianti{\textsc{chianti}}
\def\chandra{\textit{Chandra}}

\def\hii{H~\textsc{ii}}
\def\ovi{O~\textsc{vi}}
\begin{document}

\title[Missing Energy in Massive Star Clusters] {Gone With the Wind: Where is the Missing Stellar Wind Energy from Massive Star Clusters?}
%\title{Gone With the Wind: Where is the Missing Stellar Wind Energy from Massive Star Clusters?} %ApJ
\author[A.L. Rosen et al.]{Anna L. Rosen,$^1$\textsuperscript{\thanks{E-mail: alrosen@ucsc.edu}} Laura A. Lopez$,^{2} \textsuperscript{\thanks{NASA Einstein Fellow, Pappalardo Fellow in Physics}}$ Mark R. Krumholz,$^1$ Enrico Ramirez-Ruiz$^1$ \\
$^1$Department of Astronomy \& Astrophysics, University of California, Santa Cruz, CA 95064 USA \\
$^2$MIT-Kavli Institute for Astrophysics and Space Research, 77 Massachusetts Avenue, 37-664H, Cambridge MA 02139, USA\\
}

%\author{Anna L. Rosen\altaffilmark{1}, Mark R. Krumholz\altaffilmark{1}, Enrico Ramirez-Ruiz\altaffilmark{1}, Laura A. Lopez\altaffilmark{2,3,4} }
%\altaffiltext{1}{Department of Astronomy and Astrophysics, University of California Santa Cruz, 211 Interdisciplinary Sciences Building, 1156 High Street, Santa Cruz, CA 95064, USA; alrosen@ucsc.edu}
%\altaffiltext{2}{MIT-Kavli Institute for Astrophysics and Space Research, 77 Massachusetts Avenue, 37-664H, Cambridge MA 02139, USA}
%\altaffiltext{3}{NASA Einstein Fellow}
%\altaffiltext{4}{Pappalardo Fellow in Physics}

\date{Submitted 2014 May 06}

\maketitle

\begin{abstract}
Star clusters larger than $\sim 10^{3}$ $M_\odot$ contain multiple hot stars that launch fast stellar winds. The integrated kinetic energy carried by these winds is comparable to that delivered by supernova explosions, suggesting that at early times winds could be an important form of feedback on the surrounding cold material from which the star cluster formed. However, the interaction of these winds with the surrounding clumpy, turbulent, cold gas is complex and poorly understood. Here we investigate this problem via an accounting exercise: we use empirically determined properties of four well-studied massive star clusters to determine where the energy injected by stellar winds ultimately ends up. We consider a range of kinetic energy loss channels, including radiative cooling, mechanical work on the cold interstellar medium, thermal conduction, heating of dust via collisions by the hot gas, and bulk advection of thermal energy by the hot gas. We show that, for at least some of the clusters, none of these channels can account for more than a small fraction of the injected energy. We suggest that turbulent mixing at the hot-cold interface or physical leakage of the hot gas from the \hii\ region can efficiently remove the kinetic energy injected by the massive stars in young star clusters. Even for the clusters where we are able to account for all the injected kinetic energy, we show that our accounting sets strong constraints on the importance of stellar winds as a mechanism for feedback on the cold interstellar medium.\\
\end{abstract}


\section{Introduction}
Massive star clusters (MSCs; $M_{\rm \star} \ga 10^3$ $\rm M_\odot$) are born in the dense regions of giant molecular clouds (GMCs). The resulting injection of energy and momentum by the young stars (i.e., stellar feedback) terminates star formation and expels gas from the MSC. These feedback processes are likely responsible for the low star formation efficiencies observed in GMCs \citep{mm00, krumholz07}, they control the dynamics of the \hii\ regions surrounding these young clusters \citep{km09, lalopez11, lalopez13}, and they may be responsible for the dissolution and ultimate disruption of their host molecular clouds. For recent reviews, see \citet{krumholz14a} and \citet{krumholz14c}. 

Newborn stars in these clusters dramatically affect the surrounding interstellar medium (ISM) via photoionization flows (e.g., \citealt{deb13}), direct stellar radiation pressure (e.g., \citealt{km09, fkm10, mqt10}), dust-reprocessed radiation pressure (e.g., \citealt{thompson05, mqt10}), protostellar outflows (e.g., \citealt{cunningham11}), and hot gas shock-heated by stellar winds (e.g., \citealt{castor75, weaver77, canto00, stevens03, hcm09}) and supernovae (SNe; e.g., \citealt{mckee77, chevalier85}). The expansion of a cool, dense shell of interstellar material surrounding the \hii\ region is driven by these processes \citep{castor75, weaver77, mckee77, km09}. 

It is still uncertain which of these processes dominate stellar feedback and thus drive the subsequent expansion of the \hii\ region shell. Recent studies of \hii\ regions which host MSCs suggest that the thermal pressure of the warm ($\sim10^{4}$ K) ionized gas dominates \hii\ region dynamics at late times, while radiation pressure may dominate during \hii\ regions' infancy \citep[for $R_{\rm HII} \la 1-100$ pc;][]{km09, fkm10, lalopez11, lalopez13}. However, the importance of stellar winds remains uncertain \citep{lalopez11, pellegrini11, rp13, lalopez13}. The total energy injected by stellar winds is quite large for MSCs \citep{kudritzki99, repolust04}, but whether this energy contributes significantly to the dynamics of the shell, or if most of it escapes or is radiated away, remains unresolved. 

Massive stars have mass-loss rates on the order of $\dot{M}_{\rm w} \sim 10^{-6} \; \rm{M_{\odot} \; yr^{-1}}$ \citep{repolust04}. This mass escapes the stellar surface at velocities of $v_{\rm w} \sim1000 \; \rm{km \; s^{-1}}$ \citep{leitherer92}, resulting in an energy injection rate of $(1/2) \dot{M}_{\rm w} v^2_{\rm w} \sim 100 \; L_{\rm \odot}$ per massive star. In a MSC, these fast stellar winds collide with the winds of nearby stars, producing multiple shocks with complex morphologies. The hot, collective stellar wind will then produce an outward-directed ``cluster wind" as the hot gas adiabatically expands and ultimately leaves the cluster \citep{canto00, stevens03}. The resulting ``super-bubble" \citep{bruhweiler80} fills the surrounding \hii\ region with hot, shocked stellar wind material at temperatures of $\sim10^7$ K, and produces diffuse X-ray emission. This X-ray emission has been detected from numerous MSCs in the Milky Way (MW; \citealt{moffat02, townsley03, townsley06, townsley11c}) and the Large Magellenic Cloud (LMC; \citealt{lalopez11, lalopez13}). 

While the source of the X-ray emission is well understood, its luminosity is not: the X-ray luminosity of \hii\ regions associated with MSCs is less than expected if the kinetic energy supplied by winds was retained within the super-bubbles. 
There are several competing theoretical models to account for the X-ray luminosity in bubbles around MSCs, and these models imply different dynamical importance for stellar winds (and SNe, which also produce shocked hot gas). \citet{castor75} and \citet{weaver77} assume that the hot gas in the bubble is fully confined by a cool shell expanding into a uniform density ISM, whereas  \citet{chevalier85} ignore the surrounding ISM and simply assume a steady, freely-expanding wind. Compared to observations of M17 \citep{townsley03}, the Carina Nebula \citep{townsley11a}, and 30 Doradus (30 Dor; \citealt{townsley06, lalopez11}), the models of \citet{castor75} and \citet{weaver77} over-predict the observed X-ray luminosity, while that of \citet{chevalier85} under-predicts it \citep{dunne03, hcm09, lalopez11}. This result led \citet{hcm09} to introduce an intermediate model, where the hot gas expands into a non-uniform ISM, producing a ``porous" shell from which the hot gas can leak. This model is capable of fitting the observed X-ray luminosities, but the porosity is treated as a free parameter, not independently predicted.

The model of \citeauthor{hcm09} assumes the low X-ray luminosities are caused by hot gas leakage, but there are several other channels by which the wind energy can be lost. One possibility is that hot and cold gas may also exchange energy via laminar or turbulent conduction; the latter process occurs when turbulence at the hot-cold interface produces small-scale mixing of hot and cold gas, greatly accelerating conductive transfer between the two \citep{mckee84, strickland98, nakamura06}. A closely related possibility is that turbulence at the hot-cold interface mixes dust grains into the hot gas, or that dust is produced \textit{in situ} within the super-bubble by evolved stars. Dust grains immersed in hot gas will eventually be destroyed by sputtering, but they will absorb thermal energy and radiate it in the IR before that. A final possibility is that the hot gas can transfer energy to the ISM by doing mechanical work on the cold, dense shell that bounds the \hii\ region, leading to either coherent or turbulent motions \citep{breitschwerdt88}.

As previously mentioned, the effect of leakage is intimately tied to the question of stellar wind feedback in MSC formation. The contribution of stellar winds to \hii\ region dynamics, and their importance as a mechanism for limiting star formation efficiencies, depend on how much of the stellar wind energy is used to do work on the cold ISM, and how much energy is transferred to other channels.  In this paper, we attempt to determine this division of energy through observations. We examine four well-studied \hii\ regions (30 Doradus, Carina, NGC~3603, and M17), and we evaluate how the energy stored in the hot gas is lost via these different mechanisms using X-ray observations coupled to ancillary radio data. By conducting this energy census, we estimate the wind energy which leaks from the \hii\ shells, and we explore the implications regarding the role of stellar winds in regulating star formation in MSCs. This paper is organized as follows: Section \ref{sec:theory} presents the theoretical framework by reviewing the many different avenues the hot X-ray emitting gas can lose energy. Section \ref{sec:sample} discusses our source selection criteria and describes our resulting MSC sample. We present the results of our analysis of these regions in Section \ref{sec:results}, and discuss their implications in Section \ref{sec:discussion}. Finally, we summarize and conclude our analysis in Section \ref{sec:conclusion}. 

\section{Theory \& Background: Energy Budgets}
\label{sec:theory}

\subsection{$L_{\rm w}$: Energy Injection by Stellar Winds}

Consider an idealized, simple spherical \hii\ region with a MSC at its center, injecting wind energy at a rate of 
\begin{equation}
L_{\rm{w}} = \sum\limits_{i=1}^N \frac{1}{2} \dot{M}_{{\rm w,}i} v^2_{{\rm w,}i} 
\end{equation}
\noindent
where $\dot{M}_{{\rm w,}\,i}$ and $ v_{{\rm w,}\,i}$ are the mass-loss rate and wind velocity for star $i$, and $N$ is the total number of massive stars in the MSC. For typical MSCs in the MW and LMC, $L_{\rm w}$ has values of $\sim 10^{37}-10^{39} \; \rm{erg\,s^{-1}}$ \citep{crowther98, dunne03, smith06, doran13_2}. The region has a radius $R$, and is bounded by a shell of cold, dense material expanding at velocity $v_{\rm sh}$. It is filled with hot gas with temperature $T$ and electron number density $n_{\rm X}$. We assume that its filling factor is unity to assess the global dynamical effect of the hot gas on the shell \citep{lalopez11}. This picture is an oversimplification of the structure of a real H~\textsc{ii} region around a MSC. However, as we shall see below, the missing wind energy is so large that even this simplified model is able to provide meaningful constraints on the missing kinetic energy carried by stellar winds.

For the purposes of this study, we only consider the energy injected by stellar winds and ignore the contribution by SNe. This assumption is reasonable for the following reasons. First, all of the \hii\ regions considered in this paper have young MSCs (of ages $\sim$1--3 Myr) so that SNe have not occurred yet. Second, the mechanical energy of one SN ($\sim10^{51}$ erg) is comparable to the amount injected by winds over a single massive star's lifetime \citep{castor75}. As a result, ignoring SNe amounts to, at most, a factor of 2 error in the total kinetic energy injection. Finally, as we demonstrate later, we cannot account for the total energy injection by winds alone. Including the heating contribution from SNe would only strengthen our conclusions. Therefore, by using the observed X-ray, stellar population, and kinematic properties of several \hii\ regions, we examine the possible avenues that the hot gas can transfer energy in these \hii\ regions. From our analysis we determine which processes, if any, can account for the missing kinetic energy. The remainder of this section discusses the various energy sinks for the hot gas. 

\begin{figure}
\includegraphics[trim=0cm 0cm 0cm 0cm,clip,width=0.45\textwidth]{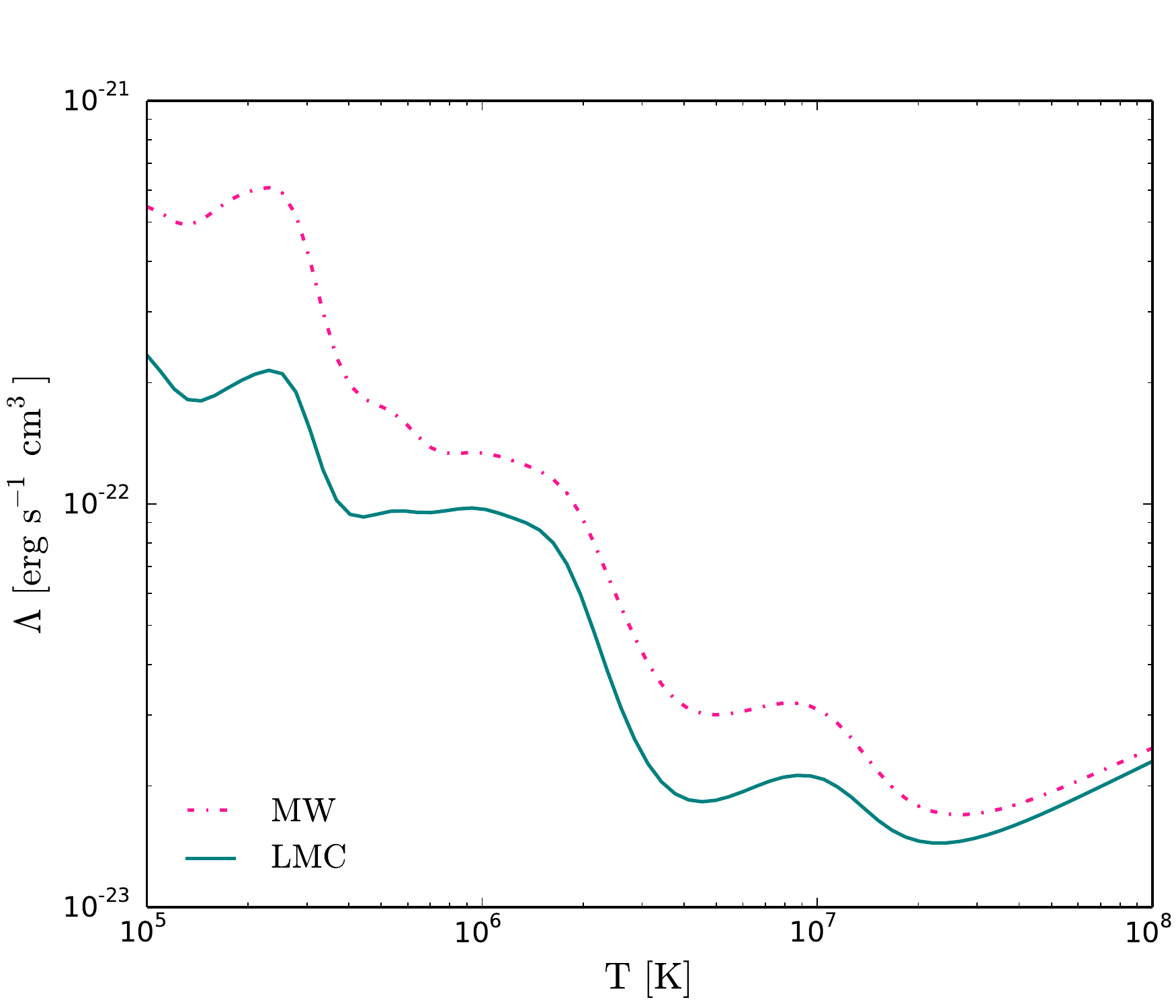}%{cooling_fcn.pdf}
\centering
\caption{
Radiative cooling functions from \chianti\ for MW ($Z=Z_{\rm \odot}$; pink dash-dot line) and LMC ($Z=0.5 \; Z_{\rm \odot}$; teal solid line) metallicities assuming that the hot gas is in CIE.
}
\label{fig:lambda}
\end{figure}

\subsection{$L_{\rm{cool}}$: Radiative Cooling of the Hot Gas}

The energy injected by winds can be lost via several channels, the first of which is radiative cooling. The hot gas cools primarily via thermal bremsstrahlung and metal-line cooling. An optically thin ``parcel" of hot gas with volume $dV$ and electron and ion number densities of $n_{\rm{X}}$ and $n_{\rm{i}}$, respectively, loses energy via radiation at a rate
 \begin{equation}
dQ = -n_{\rm{X}} n_{\rm{i}} \Lambda(T,x_{\rm i}, Z) dV dt,
\end{equation}
\noindent
where $\Lambda(T,  x_{\rm{i}}, Z)$ is the radiative cooling function (with units of $\rm{ergs \; s^{-1}\; cm^3}$), which depends on the temperature $T$, ionization state $x_{\rm i}$, and metallicity $Z$ of the hot gas. For a fully ionized plasma of Solar composition, $n_{\rm{i}}=0.9 \, n_{\rm X}$. Since $n_{\rm X}$ is dominated by the free electrons liberated from H and He, the ratio of $n_{\rm i}/n_X$ is nearly identical for the LMC and MW sources. For a low-density, optically thin plasma, $\Lambda(T, x_{\rm{i}},Z)$ is independent of density. We use \chianti\  \citep{dere97} to calculate $\Lambda(T, x_{\rm{i}}, Z)$ for MW and LMC metallicities, as shown in Figure~\ref{fig:lambda} \citep{grevesse98, russell92}. The ionization state is determined by assuming the plasma is in collisional ionization equilibrium (CIE; we discuss deviations from CIE in Section~\ref{ssec:cie}), and that charge exchange, radiative recombination, and dielectronic recombination are the only processes altering the ionization balance \citep{draine11}. In this case, the ionization fractions, $x_{\rm i}$, depend only on the gas temperature, and hence $\Lambda$ only depends on $T$ and $Z$. Under these assumptions, the total energy loss rate via cooling is
\begin{equation}
\label{eq:Lcool}
L_{\rm{cool}} = 0.9 n^2_{\rm{x}} \Lambda(T, Z) V,
\end{equation}
where $V$ is the \hii\ region volume. For typical shocked gas temperatures of \hii\ regions ($\sim10^{7}$ K), most of the photons produced by this cooling have $\sim$ keV energies, and thus are observable with X-ray telescopes, such as the \chandra\ X-ray Observatory.

We can use this result to place a constraint on the number density of the hot gas, since $L_{\rm cool} \propto n_{\rm X}^2$. If we assume that radiative cooling is the sole mechanism responsible for removing the kinetic energy injected by winds (i.e., $L_{\rm cool} = L_{\rm w}$), then the electron density of the hot gas is
\begin{equation}
\label{eq:ncool}
n_{\rm cool} = \sqrt{\frac{L_{\rm{w}}}{0.9\Lambda(T)V}}.
\end{equation}

\begin{figure*}[!t]
\centerline{\includegraphics[trim=0cm 0cm 0cm 0cm,clip,width=0.65\textwidth] {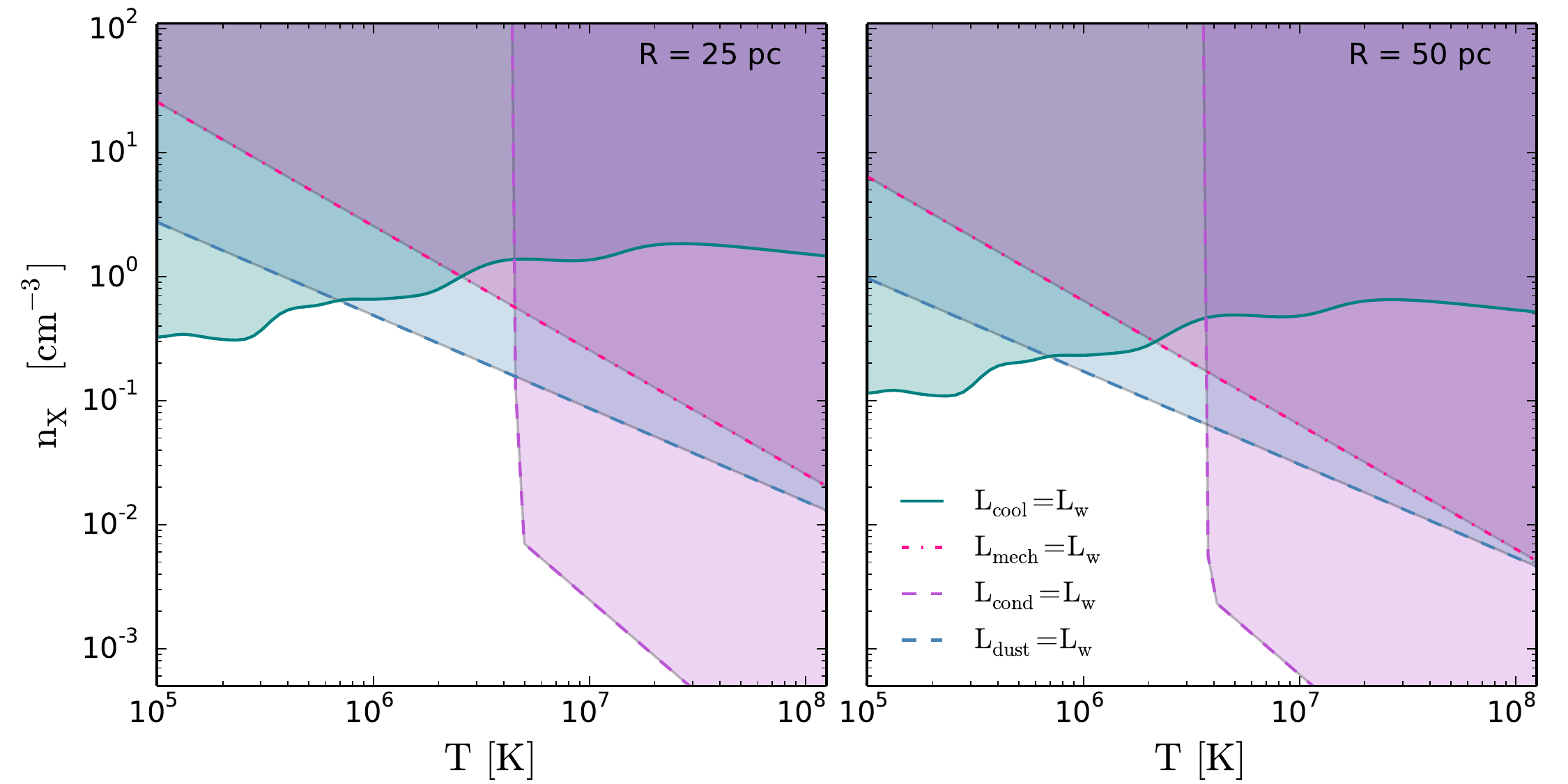}}%{nvsT_2ex.pdf}}
\caption{
\label{fig:nvsT_ex}
Allowed number densities and temperatures (white regions) for the plasma filling simulated \hii\ regions with radii of 25 pc (left panel) and 50 pc (right panel), respectively. We take $L_{\rm w}=10^{38} \; \rm{ergs \, s^{-1}}$ and $v_{\rm sh} = 20  \; \rm{km \, s^{-1}}$. Curves denote the loci in the $T-n_{\rm X}$ plane where each of the energy sinks discussed in Section \ref{sec:theory} are capable of removing all of the energy injected by winds. Shaded regions denote values of $n_{\rm X}$ and $T$ that are disallowed because the energy loss rate exceeds the energy input rate.
}
\end{figure*}

To illustrate our calculation, we consider two example \hii\ regions with radii of 25 pc and 50 pc, respectively. Figure \ref{fig:nvsT_ex} shows the value of $n_{\rm cool}$ versus temperature (solid teal line) for these two example regions, using a kinetic energy input rate of $ L_{\rm w} = 10^{38} \; \rm{erg \; s^{-1}}$. Energy conservation requires that the hot gas density and temperature must be at or below the $n_{\rm cool}$ versus $T$ curve, since the gas cannot radiate more energy than is injected by stellar winds. The shaded region above the $n_{\rm cool}$ line denotes the disallowed region and highlights the  combinations of thermodynamic properties  that violate energy conservation.

\subsection{$L_{\rm{mech}}$: Mechanical Work on the Dense Shell}

The second channel by which the hot gas can transfer energy is by doing work on the cold, dense shell that bounds the \hii\ region. Stellar feedback from the massive stars will generate a cool, dense shell of interstellar material surrounding the \hii\ region. This bubble will expand at a velocity $v_{\rm sh}$.  The hot gas, whose thermal pressure is given by $P_{\rm X}=1.9 n_{\rm X}k_{\rm B} T$, will push on the shell and, as a result, will do work $W$ at a rate $dW/dt = 4\pi R^2 P_{\rm X} v_{\rm sh }$. The factor of 1.9 arises from the fact that the total number density (e.g., $n =  n_{\rm i} + n_{\rm X}$) contributes to the pressure, and we have assumed that $n_{\rm X} = 0.9 n_{\rm i}$. We note that $P_{\rm X}=\frac{2}{3}u_{\rm X}$ where  $u_{\rm X}$ is the energy density of the hot gas and  $4\pi R^2 v_{\rm sh}$ is the rate that the volume increases as the \hii\ region expands. Under these assumptions, the rate at which the hot gas does work on the bubble shell is 
\begin{equation}
\label{eq:Lmech}
L_{\rm{mech}} = 7.6\pi R^2 v_{\rm{sh}} n_{\rm{X}} k_{\rm B} T.
\end{equation}

As with radiative cooling, we can obtain an upper limit on the electron density by considering the highest value that it could have without the resulting work exceeding the available kinetic energy supply provided by the winds (i.e., $L_{\rm mech}=L_{\rm w}$). We find that the maximum allowed number density of the hot gas is given by\begin{equation}
\label{eq:nmech}
n_{\rm{mech}} = \frac{L_{\rm{w}}}{7.6 \pi R^2 v_{\rm{sh}} k_{\rm{B}} T},
\end{equation}
which is also shown in Figure \ref{fig:nvsT_ex} (dot-dashed pink line). For this example, we take $v_{\rm sh}=20 \; \rm{km \; s^{-1}}$ which is typical for young \hii\ regions (e.g., see Table \ref{tab:HIIregions}). Similarly, since $L_{\rm mech}$ increases with increasing $n_{\rm X}$, the hot gas is only allowed to have number densities less than or equal to $n_{\rm mech}$. This result places another constraint on $n_{\rm X}$, and produces another disallowed region in the $n_{\rm X}  - T$ plane. 

We note that for temperatures below $\sim 10^6$ K, all wind energy is lost via radiation. At temperatures above a $\sim$few $\times 10^{6}$ K, mechanical work is more effective than cooling at removing the wind energy. This transition is easily discerned by calculating the ratio
 \begin{equation}
 \label{eq:lc_lm}
 \frac{L_{\rm cool}}{L_{\rm mech}} = 0.16 \frac{\Lambda(T,Z) \Sigma_{\rm X}}{k_{\rm B} T v_{\rm sh}},
 \end{equation}
 where $\Sigma_{\rm X}=n_{\rm X}R$ is the surface density of the hot gas. This ratio is less than unity for temperatures where $n_{\rm mech} < n_{\rm cool}$.
 
\subsection{$L_{\rm{cond}}$: Thermal Conduction}\label{sec:thcon} 

Conduction is a third possible kinetic energy sink. In the absence of magnetic fields, thermal conduction by the hot electrons can be an efficient energy loss mechanism at the inner edge of the cool bubble shell, since the conductive heat flux from a fully ionized plasma depends sensitively  on  temperature \citep[$\propto T^{7/2}$,][]{spitzer62}. This process creates a region of intermediate temperature gas ($T\sim 10^5$ K) between the hot bubble interior and the cold shell, and this region will shed energy rapidly via metal line cooling in the far-UV. This light would be extremely difficult to detect observationally, due to the high opacity of the ISM at these wavelengths and the even greater opacity of the Earth's atmosphere.

For classical conductivity, the heat flux is $q_{\rm{c}}=-\kappa_{\rm{c}} \bigtriangledown T$, where
\begin{equation}
\label{eq:kc}
\kappa_{\rm{c}} = 0.87 \frac{k_{\rm B}^{7/2}T^{5/2}}{m^{1/2}_{\rm{e}} e^4 \ln \Lambda_{\rm{C}}}
\end{equation}
is the thermal conductivity of the hot electrons with temperature $T$, and
\begin{equation}
\label{eq:lc}
\ln{\Lambda_{\rm{C}}} = 29.7 + \ln{\left(n_{\rm X}^{-1/2} \frac{T}{10^6 \rm{\,K}}\right)}
\end{equation}
\noindent
is the Coulomb logarithm for $T>4.2 \times 10^5$~K \citep{spitzer62, cowie77, draine11}. \citet{cowie77}  find that when the electron mean free path becomes comparable to or greater than the temperature scale height, $T/|\bigtriangledown T|$, the heat flux saturates and takes on the value 
\begin{equation}
\label{eq:qs}
q_{\rm{s}} = 0.4 \left(\frac{2k_{\rm B}T}{\pi m_{\rm{e}}}\right)^{1/2} n_{\rm{X}} k_{\rm B} T.
\end{equation}
\noindent
The total energy loss rate due to conduction for an \hii\ region with radius $R$ filled with hot gas at a temperature $T$ is therefore
\begin{equation}
\label{eq:Lcond}
L_{\rm cond} = 4\pi R^2 \min(\kappa_c |\nabla T|, q_s).
\end{equation}
We further assume that $|\nabla T| \sim T/R$, which is true at the order of magnitude level.

If thermal conduction is responsible for removing the bulk of the energy injected by stellar winds ($L_{\rm cond} = L_{\rm w}$), then the required number density of the hot gas is
 \begin{equation}
 \label{eq:ncondc}
 n_{\rm cond} = \left(\frac{T}{10^6\rm{K}}\right)^2 \exp{\left(59.4-6.96\pi \frac{k_{\rm B}^{7/2} T^{7/2} R}{m^{1/2}_{\rm e} e^4 L_{\rm w}}\right)},
 \end{equation}
assuming that the heat flux  is not saturated. This result follows from Equations \ref{eq:kc}, \ref{eq:lc}, and \ref{eq:Lcond}. However, if the temperature is large enough such that the conductive heat-flux becomes saturated, then the number density required for conduction to dominate the energy loss is
 \begin{equation}
 \label{eq:nconds}
 n_{\rm cond} =  \left( \frac{m_{\rm e}}{2 \pi}\right)^{1/2} \frac{L_{\rm w}}{1.6 R^2 k_{\rm B}^{3/2} T^{3/2}},
 \end{equation}
\noindent
which follows from Equations \ref{eq:qs} and \ref{eq:Lcond}. These results are also shown in Figure \ref{fig:nvsT_ex} (purple dashed line). The shaded region to the right of this curve denotes the forbidden region where conductive losses exceed wind energy input. 

Finally, we note that Equation~\ref{eq:Lcond} is almost certainly a large overestimate of the true conductivity, because a non-radial magnetic field, even a dynamically sub-dominant one, will greatly reduce the heat flux \citep{soker94}. We address the effects of magnetic fields on conduction in more detail in Section \ref{ssec:magneticfields}. If the conductive heat flux is less than Equation \ref{eq:Lcond}, then $n_{\rm cond}$ will shift to higher temperatures, thereby reducing the size of the forbidden region. 

\subsection{$L_{\rm dust}$: Collisional Heating of Dust Grains}

The next energy sink we consider is the transfer of thermal energy from the hot gas to dust grains via collisions, followed by thermal radiation from the grains. The molecular clouds out of which MSCs form are dusty, and this dust can mix with the hot gas in two ways. First, the dust in ISM material can mix with the shocked wind material. Second, the expanding shell around the \hii\ region will become corrugated with instabilities \citep{strickland98}, and the resulting turbulence at the hot-cold interface can mix dust grains into the hot gas. One final process by which dust can be supplied to and mixed with the hot gas is independent of the molecular cloud material: \textit{in situ} formation of dust surrounding evolved stars such as red supergiants in the young MSC \citep{levesque10}. Regardless of its source, dust grains immersed in hot gas will eventually be destroyed by sputtering, but they will absorb thermal energy via inelastic collisions and radiate it in the IR before that.  

The importance of these processes depends on how well the dust is mixed with the hot gas and on how the sputtering and destruction time scales compare \citep{smith96,draine11}. These parameters depend on the properties of the dust grains and on the density and temperature of gas in the turbulent mixing layer. We address this question in detail in Section \ref{ssec:dustlifetime}, but to be conservative we perform the calculation assuming that dust is able to survive in the hot gas with the same abundance as in the cold gas. Under this assumption, the gas-dust energy exchange rate by collisions with dust grains per unit volume of the hot gas is given by
\begin{equation}
\label{eq:psi_gd}
\Lambda_{\rm gd} = n_{\rm X} n_{\rm d} \sigma_{\rm d} \left( \frac{8k_{\rm B} T}{\pi m_{\rm e}} \right)^{1/2} \bar{\alpha}_{T} \left( 2 k_{\rm B} T_{\rm d} - 2 k_{\rm B} T \right) 
\end{equation}
\noindent
where $n_{\rm d}$ is the dust grain number density, $T_{\rm d}$ is the dust temperature, $T$ is the hot gas temperature, $\sigma_{\rm d}=\pi a^2 \gamma_e$ is the dust cross section where the factor $\gamma_e$ allows for Coulomb focusing/repulsion of the hot electrons. Here  $\bar{\alpha}_T$ is the averaged accommodation coefficient for an astrophysical mixture of gases which describes the fraction of kinetic energy of the impacting electron may be converted to heat \citep{burke83, draine11, krumholz13}. For dust grains immersed in hot gas with temperatures greater than $\sim10^6$~K, the electric potential is much less than the thermal energy of the impacting electrons, and thus the dust grain can be treated as neutral \citep[i.e., $\gamma_{\rm e}=1$;][]{dwek87}. We assume that the accommodation coefficient is equal for both H atoms and the hot electrons, with a value $\bar{\alpha}_{\rm T} = 0.3$ \citep{burke83, dwek87, krumholz11}. For canonical values of the dust-to-gas mass ratio and the dust grain cross section, assuming that the total surface area of grains is proportional to the metal abundance, we find that the total energy exchange rate from the hot gas to the dust is
\begin{equation}
\label{eq:Ldust}
L_{\rm dust} = \alpha_{\rm dg,\,e} n_{\rm X}^2 V T^{3/2}, 
\end{equation}
\noindent
where 
\begin{equation}
\alpha_{\rm dg, \, e}=2.20\times10^{-31} \frac{Z}{Z_{\rm \odot}}\rm{erg \; cm^{3} K^{-3/2} s^{-1}}
\end{equation}
\noindent
is the grain-gas coupling parameter, which is proportional to $n_{\rm X}/m_{\rm e}^{1/2}$ \citep{krumholz11}.

If heat exchange from the gas to the dust is primarily responsible for removing the bulk of the energy injected by winds, then the number density of the hot gas is
\begin{equation}
\label{eq:ndust}
n_{\rm dust} = \sqrt{\frac{L_{\rm w}}{\alpha_{\rm dg, \, e} V T^{3/2}}}.
\end{equation}
\noindent
These results are also shown in Figure \ref{fig:nvsT_ex} (dashed blue line). Again, the shaded region above this curve denotes the forbidden region within which the energy loss rate exceeds the injection rate.

Finally, we warn the reader that Equation~\ref{eq:Ldust} is likely an overestimate of the true energy transfer rate of the hot gas colliding with dust. The value of $\alpha_{\rm dg, \, e}$ is dependent on the adopted dust-to-gas ratio. Here, we have assumed that the dust-to-gas ratio for the hot gas is the same as that of the neutral ISM, where dust is perfectly mixed with the gas. The true dust-to-gas ratio in the hot gas is likely to be smaller, and the value will depend on the competition between turbulent mixing at the hot-cold interface and the sputtering of grains in the high temperature medium.

\subsection{$L_{\rm{leak}}$: Physical Leakage of the Hot Gas}

The final energy sink that we will calculate is that associated with bulk motion of the hot gas. The hot gas may be only partially confined by the cold gas in the shell, either because the surrounding ISM is non-uniform or because stellar feedback punches holes in the shell. In either case, the hot gas, which has a much larger sound speed than the cool gas, will flow out of the holes, expand adiabatically, and cool radiatively \citep{hcm09}. In this scenario, the energy injected by stellar winds is ultimately radiated as low-surface brightness X-ray and far-UV emission over an area much larger than the observed \hii\ region.

\citet{hcm09} define a confinement parameter $C_{\rm{f}}$  that describes the ``porosity" of the cold shell, where $C_{\rm{f}} = 1$ describes a shell with no holes and $C_{\rm{f}} = 0$ describes a completely porous shell  (i.e., no shell exists). The holes allow the hot gas to escape the \hii\ region at its sound speed, $c_{\rm{s}}$, with an energy flux given by
\begin{equation}
\label{eq:Lleak}
L_{\rm{leak}} = \left(1-C_{\rm{f}}\right) 4\pi R^2 \frac{5}{2} \rho_{\rm{h}} c^3_{\rm{s}}
\end{equation}
\noindent
where $\rho_{\rm{h}}=1.9 \mu m_{\rm{p}} n_{\rm{X}}$ is the density of the hot gas and $\mu=0.62$ assuming He is fully ionized and its mass fraction is 0.3. Note that since $L_{\rm{leak}}\propto c^3_{\rm{s}}$, a large amount of leakage can occur even if the  shell has a large covering fraction.

\subsection{Other Forms of Energy Loss}

The energy losses by the mechanisms discussed previously in this section can be estimated easily under the stated assumptions. However, other avenues of energy loss also exist, including ``turbulent conduction" and ``turbulent work." As we shall see below, the former mechanism may likely dominate the energy loss of the hot gas. Unfortunately, both channels are much harder to assess using observations, even at the order-of-magnitude level. Nonetheless, we summarize these underlying physical mechanisms here.

Turbulent conduction \citep{mckee84} describes how cold gas can mix rapidly with hot gas via Kelvin-Helmholtz instabilities that occur as hot gas flows past cold clouds, either in the hot bubble interior or as it leaks out \citep{strickland98, nakamura06}. As with the estimate provided above for thermal conduction, this mixing will lead to rapid conductive transmission of the thermal energy, producing gas at temperatures of $\sim 10^5$ K which sheds energy rapidly via metal line cooling in the far-UV.

The difference between thermal conduction and turbulent conduction is that if a turbulent mixing layer is present, then the effective area of the hot-cold interface and the sharpness of the temperature gradient can be orders of magnitude larger than the laminar estimate given by Equation~\ref{eq:Lcond}. Moreover, if the turbulent mixing layer produces mixtures of hot and cold gas on scales smaller than the electron gyro radius, then magnetic confinement of the electrons will not be able to restrict the rate of energy interchange between hot and cold gas. Thus, the presence of a turbulent mixing layer might lead to a conductive loss rate much higher than the simple laminar estimate presented in Section~\ref{sec:thcon}.

The final energy loss mechanism we consider is turbulent work, where hot gas collides with the cold ISM and does work on it, converting its thermal energy into a turbulent cascade in the cold gas \citep{breitschwerdt88}. This process leads to the formation of shocks and to the energy being radiated in the IR (if dust cooling of the cold ISM dominates) or radio (if molecular line cooling dominates). Turbulent work is related to the mechanical luminosity we have estimated above, but it would manifest as large incoherent velocities rather than the coherent expansion in the previous estimate. It is unlikely that turbulent work would dominate over mechanical work because the total amount of work done on the cold ISM depends on the total surface area of the working surface (i.e., the bubble shell). As the shell expands coherently, the work done on the shell will be much greater than that over the turbulent regions. We therefore conclude that turbulent work is not a dominant energy sink for the hot gas.

\section{H\textsc{II} Region Sample}
\label{sec:sample}

\begin{table*}
\begin{center}
%\begin{minipage}{125mm}
	\caption{Inferred number densities, luminosities, and confinement factors}
	 \label{tab:HIIregions}
	\begin{tabular}{lccccccccr}
	
	Name & $D\,\rm{[kpc]}$ & $\rm{R_{sh}\,\rm{[pc]}}$ & $v_{\rm{sh}} \,\rm{[km/s]}$ & $t_{\rm{cl}} \, \rm{[Myr]}$ & $\log L_{\rm{bol}} \,\rm{[L_{\odot}]}$ & $Q \,\rm{[10^{49} \, s^{-1}]}$ & 
					$L_{\rm w} \,\rm{[10^{37} \, erg/s]}$ & $L_{\rm x} \,\rm{[10^{35} \, erg/s]}$ & $T_{\rm X} \, [10^6\;K]$ \\
	\hline
	\hline
	30 Doradus & $50$ & $100$& $25$ & $2$ & 8.4 & 1237 & 224 & 45.0 & 7.4\\
	Carina & 2.3 & $20$ & $20$ & $3$ & 7.23 & 90 & 35.0 & 1.71 & 4.5 \tablenotemark{a}\\
	NGC 3603 & 7.0 & $21$ & $20$ & $1$ & -- & 155 & 62.0 & 2.6 & 6.2 \tablenotemark{a}\\
	M17 & 2.1 & $5.8$ & $25$ & $1$ & $6.58$ & 14.3 &$1$ & 0.2 & 5.3 \tablenotemark{a}\\
	\hline
	\end{tabular}
	
\tablenotetext{a}{Temperatures shown are surface-brightness weighted values from \citet{townsley11c}.}
\tablerefs{\textbf{30 Doradus}: \citet{lalopez11, doran13_2, lalopez13}; \textbf{Carina}: \citet{smith00, smith06, smith07, hcm09, townsley11c}, \textbf{NGC 3603}: \citet{balick80, crowther98, townsley11c}, \textbf{M17}: \citet{clayton85, dunne03, townsley03, hoffmeister08, townsley11c}}
%\end{minipage}
\end{center}
\end{table*}

\subsection{Sample Selection Criteria}
In the previous section, we have reviewed the various physical processes that can contribute to the energy accounting problem. In the following sections, we derive constraints on the effectiveness of each process in depleting the injected wind energy from a sample of well-studied \hii\ regions. We have selected this sample using several criteria. First, there must be X-ray data available to enable us to determine the physical properties of the hot gas and estimate the rate of radiative energy losses.  Second, we require radio observations that allow us to characterize the cold shells bounding the \hii\ regions, since the radius and velocity of this shell enter in our estimates for the mechanical luminosity.

Finally, we require robust observational estimates of the wind energy output by the stars. To obtain an accurate accounting of the wind energy, the spectral types of the majority of the luminous stars are necessary, so that star-by-star surface gravities and temperatures can be determined. Given these constraints, we have restricted our analysis to four well-studied \hii\ regions in the LMC and MW, which we describe briefly below. All our sources have young aged clusters ($\sim$1--3 Myr old), and thus their X-ray emission is predominantly powered by stellar winds from their MSCs.

\subsection{Individual H\textsc{II} Regions}
\subsubsection{30 Doradus}
30 Doradus (hereafter 30 Dor), located in the LMC (at a distance $D \sim50$ kpc), is the most luminous and largest \hii\ region in the Local Group, with a radius of $\sim 100$ pc and a bolometric luminosity of $\sim 2.3\times10^{8} \; L_{\rm{\odot}}$ \citep{lalopez13,doran13_2}. It is primarily powered by NGC 2070 which contains $\sim2400$ OB stars. At its center lies R136, a young ($t_{\rm{age}}\sim1-2$ Myr) dense star cluster with a stellar mass density of $5.5\times10^4 \;\rm{M_{\odot} \; pc^{-3}}$ \citep{parker93, hunter95}. The total energy input by the stellar winds is $2.2\times10^{39}\;\rm{erg \; s^{-1}}$ \citep{doran13_2}, and its bubble shell is expanding at an average speed of $\sim25 \;\rm{km\;s^{-1}}$ \citep{chu94}. The X-ray emission from 30 Dor was observed using the \chandra\ Advanced CCD Imaging Spectrometer (ACIS) for an integrated time of $\approx 94$ ks (PI: L. Townsley).  \citet{lalopez11} found that the total diffuse unabsorbed X-ray luminosity of 30 Dor in the 0.5--2 keV band is $4.5\times10^{36}~\rm{erg \; s^{-1}}$. From the X-ray spectra, \citet{lalopez11} found that the X-ray emission can be characterized by a hot plasma with a temperature of $7.4\times10^6$~K.

\subsubsection{The Carina Nebula}
The Carina Nebula, located at a distance of $D \sim2.3$~kpc, is one of the nearest regions of active massive star formation \citep{allen93, smith06}. This complex is a ``cluster of clusters", containing 8 open clusters. It hosts $\sim$70 O stars, of which 46 belong to the young star cluster Trumpler~16 \citep[$t_{\rm cl} \sim 2-3$ Myr;][]{smith06}, the home of the well-known luminous blue variable $\eta$ Carina. The total bolometric luminosity of the stars in the Carina Nebula is  $2.5\times10^{7} \; \rm{L_\odot}$, and the total energy input by stellar winds is $\sim 3.5\times 10^{38} \; \rm{erg \; s^{-1}}$, with 70\% of the energy budget coming from Trumpler~16 \citep{smith06, hcm09}. The nebula has a radius of $\sim20$ pc \citep{hcm09, townsley11c} and its outer shell is expanding at a velocity of $\sim20 \rm{\;km\;s^{-1}}$ \citep{smith00,smith07}. Carina has been studied extensively with \chandra. \citet{townsley11a} obtained a 1.2~Ms, 1.42 deg$^2$ ACIS-I mosaic of the complex to characterize its diffuse emission and to identify thousands of X-ray point sources (e.g., low-mass  pre-main sequence stars and massive stars). They found that the total integrated diffuse emission from the Carina Nebula in the 0.5--7 keV X-ray band is $1.7\times10^{35}\; \rm{erg\;s^{-1}}$. To derive the temperature of the hot gas, \citet{townsley11c} fit the X-ray spectra with three non-equilibrium ionization (NEI) plasma components. For our purposes, we assume that the observed hot gas temperature is the surface-brightness weighted value taken from \citet{townsley11c}. This yields a temperature of $4.5\times10^6$ K. 

\subsubsection{NGC 3603}
The giant \hii\ region NGC 3603, located at a distance of $D \sim7$ kpc, contains the most compact and youngest ($t_{\rm{cl}} \approx 1$ Myr) massive ``starburst" cluster located in the MW \citep[HD 97950;][]{crowther98}. With a mass density of $\sim10^5 \; \rm{M_\odot\;pc^{-3}}$,  HD 97950 is more compact than R136 \citep{hofmann95}. Assuming a distance of $D \sim8.4$ kpc \citep{goss69}, \citet{balick80} found that the radius of NGC 3603 (i.e., the region which contains 90\% of the radio flux) is 25 pc. Assuming a distance of $7$ kpc, the radius of NGC 3603 reduces to $\sim21$ pc. \citet{balick80} also studied the dynamics of NGC 3603 by measuring multiple emission lines, including H$\alpha$ and $\rm{N}~\textsc{ii}$. They found that the $\rm{N}~\textsc{ii}$ lines are double peaked and separated by $\sim20$ $\rm{km \; s^{-1}}$. Furthermore, they also measured the velocity dispersion of the H$\alpha$ turbulent line widths to be 20 $\rm{km \; s^{-1}}$. These results suggest that NGC 3603 is expanding at a rate of $\sim20$ $\rm{km \; s^{-1}}$. Performing a stellar census of the massive stars in NGC 3603, \citet{crowther98} estimate that the total mechanical energy input by stellar winds is $6.2\times10^{38}\; \rm{erg \;s^{-1}}$. \citet{smith06} suggest that the actual wind luminosity for NGC 3603 is smaller than this value because \citet{crowther98} do not consider the effect of wind clumping in their analysis. However, \citet{smith06} does not quantify what this value is so we adopt the wind luminosity estimate from \citet{crowther98} and warn the reader that this may likely be an overestimate. NGC 3603 was observed with \chandra\ with 46 ks of usable time \citep{moffat02}. \citet{townsley11c} re-analyzed these data and found 1328 X-ray point sources in the $17' \times 17'$ ACIS-I field. After removing these point sources, \citet{townsley11c} found that the diffuse emission of NGC 3603 in the 0.5--7 keV band is $2.6\times10^{35} \; \rm{erg \;s^{-1}}$. Similarly to Carina, they fit the X-ray spectra by a multiple component plasma. Taking the surface-brightness weighted average of their results, we adopt a hot gas temperature of $6.2\times10^6$ K.

\subsubsection{M17}
The emission nebula M17, located at a distance of $D \sim2.1$ kpc, is on the eastern edge of a massive molecular cloud, M17SW, and exhibits a blister-like structure with a radius of $\sim$5.8 pc \citep{townsley03}. It is powered by the open cluster NGC 6616, which consists of a ring of seven O stars $\sim0.5$ pc in diameter \citep{townsley03}. NGC 6616 is quite young, with an estimated age of $\lesssim1$ Myr \citep{hanson97}. Assuming a distance of 2.1 kpc to the nebula, \citet{hoffmeister08} found that the total bolometric luminosity of the stellar population in M17 is $3.8\times10^6 \; \rm{L_{\rm \odot}}$. \citet{dunne03} estimate that $L_{\rm{w}}\sim1\times10^{37} \;\rm{erg \; s^{-1}}$. The bubble shell of M17 is expanding at a rate of $\sim25 \; \rm{km \; s^{-1}}$\citep{clayton85}. \citet{townsley09} created a deep (total ACIS-I integration time of 320 ks) mosaic of M17. From these data, \citet{townsley11c} found that the total integrated diffuse emission from M17 in the 0.5--7 keV X-ray band is $2.0\times10^{34}\; \rm{erg\;s^{-1}}$. They modeled the X-ray spectra by a multiple component plasma, and from their model we adopt the surface-brightness weighted value of $5.3\times10^6$ K for the hot gas temperature. 

\section{Results}
\label{sec:results}

In this section, we assess which sinks are responsible for removing the wind kinetic energy injected by MSCs in our four \hii\ regions. We perform this analysis in several steps. First, in Section \ref{ssec:obsconstraints}, we constrain the actual densities and temperatures of the hot gas in our sample \hii\ regions using the available observations of their diffuse X-ray emission. Second, in Section \ref{ssec:energysinks}, we evaluate all of the sink terms discussed in Section \ref{sec:theory} to determine which of them, if any, might be responsible for removing the bulk of the wind energy. We use these calculations to evaluate the global energy budget for stellar wind energy injection in Section \ref{ssec:budget}.

\begin{figure}
\centerline{\includegraphics[trim=0cm 0cm 0cm 0cm,clip,width=0.45\textwidth]{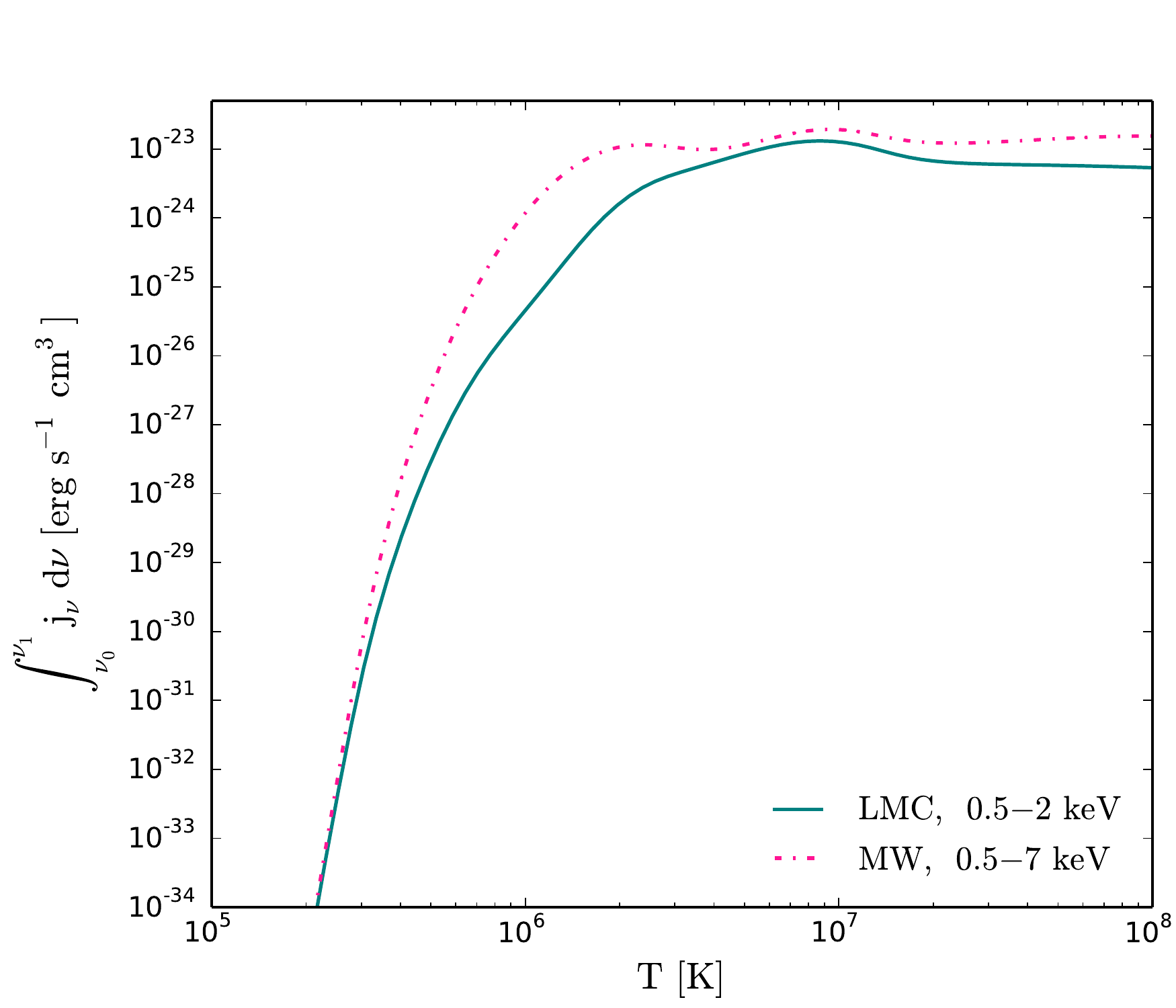}}%{emissivity_fcn.pdf}}
\caption{
\label{fig:emissivity}
Frequency-integrated emissivities from \chianti\ for MW ($Z=Z_{\rm \odot}$) and LMC ($Z=0.5 \; Z_{\rm \odot}$) metallicities assuming that the hot gas is in CIE. The LMC emissivity is integrated over the 0.5-2 keV \chandra\ band and the MW emissivity is integrated over the 0.5-7 keV \chandra\ band.
}
\end{figure}

\subsection{Observational Constraints on the Hot Gas Density and Temperature}
\label{ssec:obsconstraints}

The density and temperature of the hot gas are jointly constrained by the observed (absorption-corrected) X-ray luminosity, while the temperature is constrained by the shape of the X-ray spectrum. We focus on the former constraint first. A ``parcel" of hot gas with temperature $T$, electron number density $n_{\rm{X}}$, and volume $V$ will have an X--ray luminosity given by
\begin{equation}
L_{\rm{X,\;obs}} = 0.9 n_{\rm{X}}^2 V \int^{\nu_1}_{\nu_0} j_{\rm{\nu}}(T,Z) d\nu
\end{equation}
where $j_{\rm{\nu}}(T, Z)$  is the emissivity of the hot gas and ($\nu_0,\;\nu_1$) is the frequency band of the X-ray telescope. For our purposes, we focus on the soft (0.5--2 keV) X-ray band when available, since the luminosity at these energies originates from the diffuse structures created by the collision of stellar winds. These are brighter by an order of magnitude than the point sources \citep{townsley06}. From the literature, we only have $L_{\rm X}$ for the 0.5--2 keV band for 30 Dor \citep{lalopez11}, whereas we have $L_{\rm X}$ for the 0.5--7 keV band for the MW \hii\ regions \citep{townsley11c}. We use \chianti\ to compute $j_{\rm{\nu}}(T,Z)$ for both MW and LMC abundances \citep{russell92, grevesse98}, and we show the results in Figure \ref{fig:emissivity}. Under our simple assumption of a uniform hot gas filling the \hii\ region, we can then combine the observed luminosity with the approximate volume of the region to obtain the number density of the hot X-ray emitting gas,
\begin{equation}
\label{eq:nx}
n_{\rm{X}} = \sqrt{\frac{L_{\rm X,\;obs}}{0.9 V \int^{\nu_1}_{\nu_0} j_{\rm{\nu}}(T,Z) d\nu}}.
\end{equation}

From the X-ray data, one can also determine the temperature of the hot gas by modeling the X-ray spectrum as an absorbed hot diffuse gas. For this purpose, we adopt the surface-brightness weighted temperatures derived from the observations, as discussed in Section \ref{sec:sample}.

\begin{figure}
\centerline{\includegraphics[trim=0cm 0cm 0cm 0cm,clip,width=0.45\textwidth]{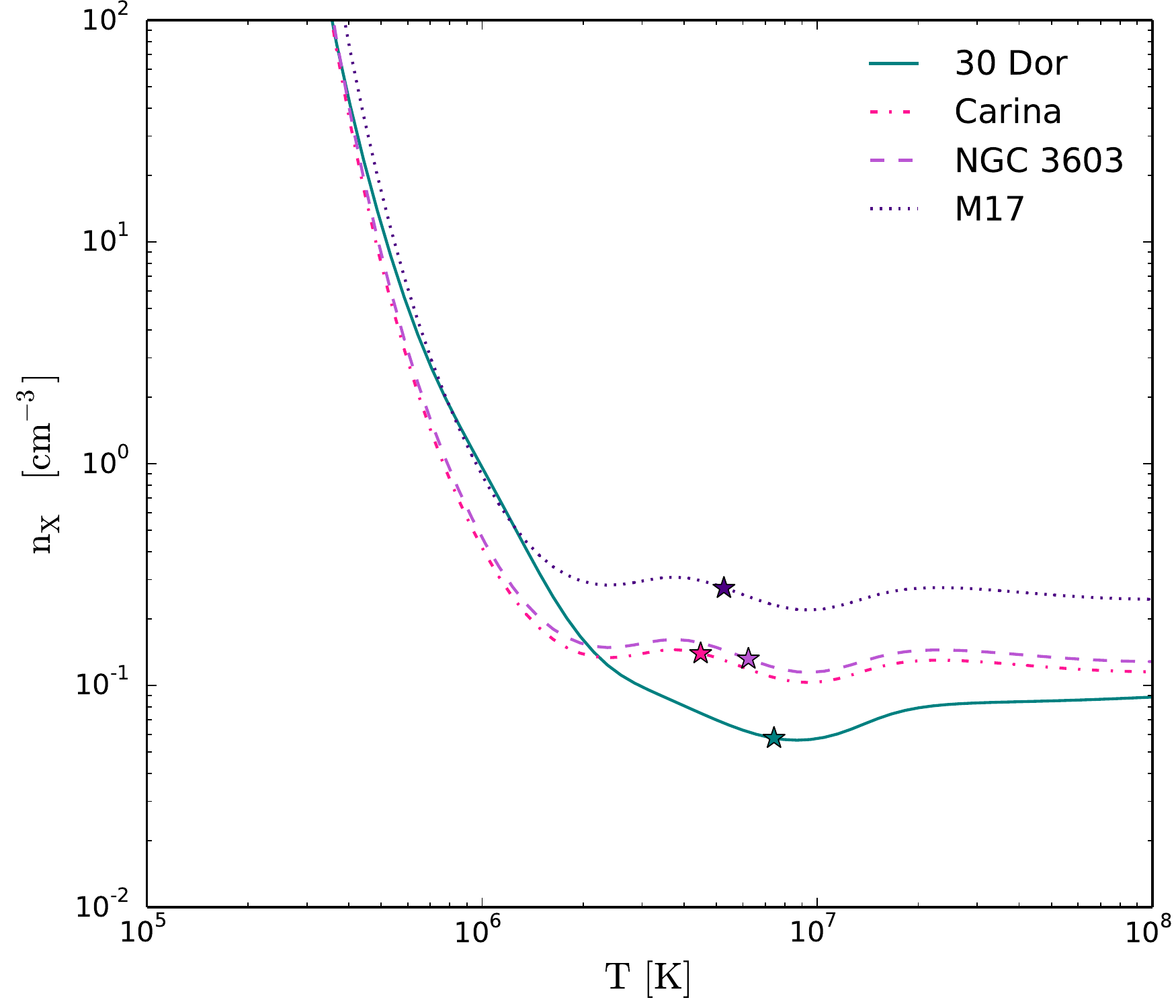}}%{nxvsT_all.pdf}}
\caption{
\label{fig:nxvsT}
Allowed hot gas number density, $n_{\rm X}$, versus temperature, $T$, constrained by X-ray observations for the 30 Dor, Carina, NGC 3603, and M17 \hii\ regions.  As can be seen, the allowed $n_{\rm X}$ for the MW \hii \ regions (Carina, NGC 3603, and M17) follow the same shape but have different offsets due to their differing $L_{\rm X, \;obs}$. The points denote the temperatures inferred by spectral fitting (see Table \ref{tab:HIIregions}).
}
\end{figure}

Figure \ref{fig:nxvsT} illustrates the locus in the $T-n_{\rm X}$ plane allowed by the observed luminosities of our sample \hii\ regions, with points marked along these curves corresponding to the temperatures inferred from the spectra. The $n_{\rm X}$ versus $T$ curves for the MW sources have the same shape because they all use the same metallicity and bandpass for $j_\nu(T,Z)$, but have different observed luminosities. The curve for 30 Dor in the LMC has a slightly different shape due to the difference in both the frequency band used for the observations and in the gas metallicity.

\subsection{Energy Sinks}
\label{ssec:energysinks}

\begin{figure*}
\centerline{\includegraphics[trim=0cm 0.5cm 0cm 0cm,clip,width=0.6\textwidth]{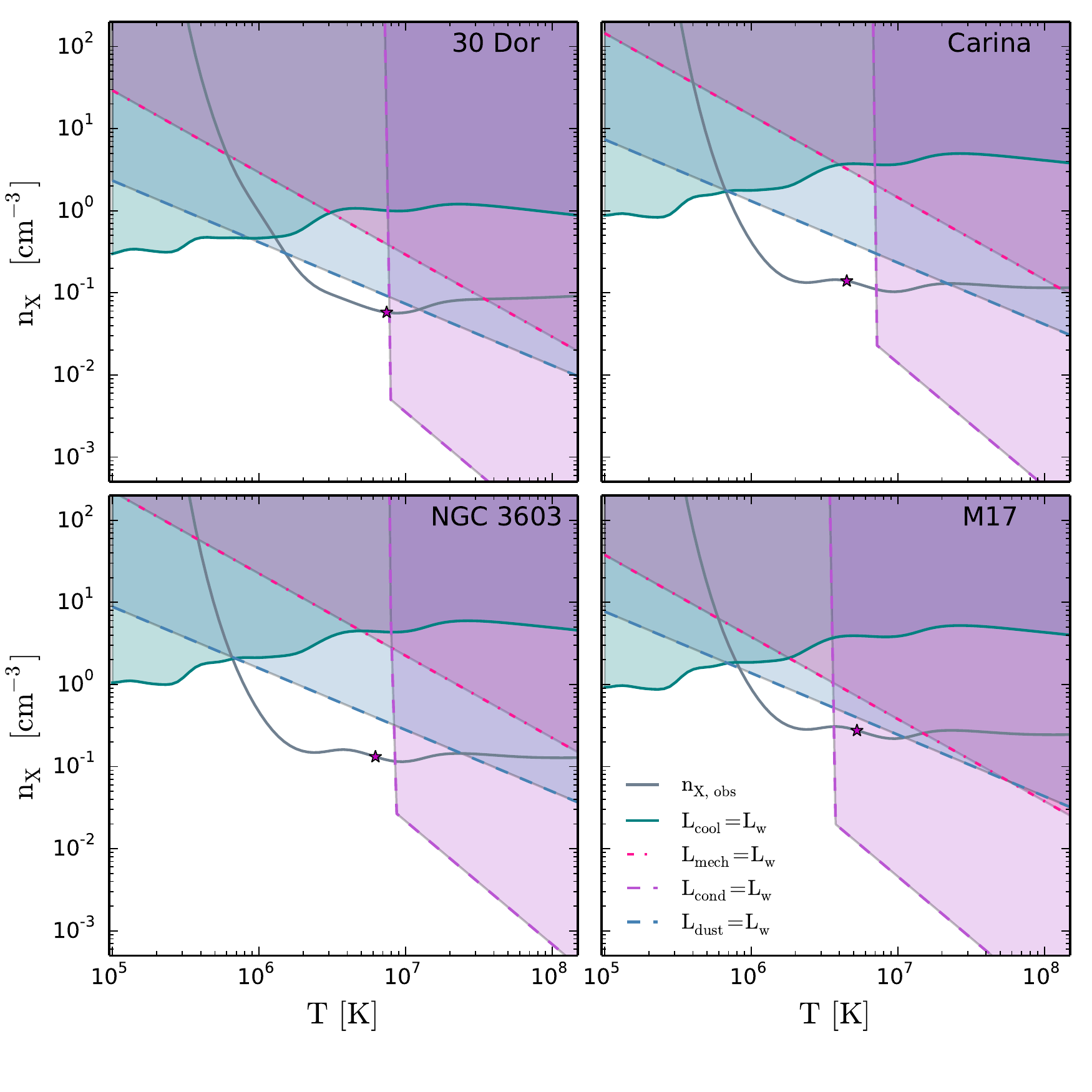}}%{AllClusterfill_nvsT.pdf}}
\caption{
\label{fig:nvsT}
Same as Figure \ref{fig:nvsT_ex} but for the \hii\ regions in our sample. The gray lines with points along them indicate the values of $n_{\rm X}$ versus $T$ constrained by the observed X-ray luminosities, with the points indicating the temperatures inferred by spectral fitting (see Table \ref{tab:HIIregions}).
}
\end{figure*}

We next estimate the energy sinks discussed in Section \ref{sec:theory} for our sample \hii\ regions, in order to produce for each one a plot of the same type as shown in Figure \ref{fig:nvsT_ex} (i.e., the loci in the $T - n_{\rm X}$ plane where each potential energy sink is capable of removing all of the kinetic energy injected by the winds). The inputs to these calculations are the observed \hii\ region properties given in Table \ref{tab:HIIregions}. We show the results of these calculations in Figure \ref{fig:nvsT}, with the curves of $n_{\rm X}$ versus $T$ inferred from the observed X-ray emission overlaid.

The $L_{\rm cool} = L_{\rm w}$ curve (i.e., Equation \ref{eq:ncool} -- the solid teal line) indicates density-temperature combinations such that all the kinetic energy injected by stellar winds is radiated away. We remind the reader that the hot gas can only lie on or below $n_{\rm{cool}}$ line in order to conserve energy (i.e., the gas can not radiate more energy than is injected into it). Clearly, the required number densities for cooling to dominate the energy loss are much larger than the number density constrained by the observed X-ray emission for all \hii\ regions in our sample for $T\gtrsim 10^6$ K.  The observationally-inferred gas temperatures are well above this limit. We conclude that radiative cooling is not an important energy sink, consistent with previous results \citep{dunne03, lalopez11, townsley11c}.

Next, we consider the $L_{\rm mech} = L_{\rm w}$ curve (i.e., Equation~\ref{eq:nmech} -- the dot-dashed pink line), the locus of density-temperature combinations for which mechanical work on the dense shell removes the bulk of the wind energy. We find that for temperatures of $\gtrsim 1-2 \times 10^6$ K, mechanical work becomes more efficient at removing energy since the hot gas pressure increases with temperature and $L_{\rm mech} \propto n_{\rm X}$. Thus, the $L_{\rm mech} = L_{\rm w}$ curve requires lower number densities than radiative cooling to remove the wind energy. However, we find that $n_{\rm mech}$ is still larger than the number density constrained by the observed X-ray emission for all \hii\ regions unless the hot gas temperature exceeds $\sim 0.2-1 \times 10^8$ K. None of the \hii\ regions in our sample are in this temperature range. Thus, we conclude that mechanical work on the bubble shell is not responsible for removing the bulk of the wind energy.

The next energy sink we consider is thermal conduction. The $L_{\rm cond} = L_{\rm w}$ curve (i.e., equations \ref{eq:ncondc} and \ref{eq:nconds} -- the dashed purple line) is nearly vertical at low temperatures because $L_{\rm cond}$ depends only weakly on density (e.g., equation \ref{eq:lc}) in the unsaturated regime. Only when the heat flux reaches the saturated value does the conductive luminosity exhibit any significant density dependence. We find that the observationally-constrained number densities and temperatures do lie in the region where conduction is capable of removing the bulk of the wind energy for 30 Dor and M17. However, we remind the reader that our estimate of the conductive heat flux is almost certainly a sizable overestimate, as we have entirely neglected the effects of magnetic fields. Thus, our results show that for densities and temperatures consistent with observations, thermal conduction can be an important energy sink for stellar wind energy as long as it is not significantly inhibited by magnetic fields.

Lastly, we consider the energy transfer of the hot gas to dust via collisions. The $L_{\rm dust} = L_{\rm w}$  curve (i.e., Equation \ref{eq:ndust} -- the dashed blue line) indicates the density-temperature combinations at which all of the energy injected by stellar winds is transferred to the dust via collisions with the hot gas. We find that the heating of dust is more effective at removing energy from the hot gas than cooling and mechanical work for $T \gtrsim 10^6$ K. We also find that the $L_{\rm dust} = L_{\rm w}$ curve is quite close to the observational constraint line $n_{\rm X,obs}$ for temperatures consistent with the observed spectrum in 30 Dor and M17. This result suggests that dust heating could be a significant energy sink for 30 Dor and M17, but probably not in NGC 3603 and Carina. However, as with conduction, our energy loss estimates for dust heating are likely to be large overestimates, since they assume that the dust content in the hot gas matches that in the cool ISM.

We defer a calculation of the rate of energy leakage via bulk motion to the following Section since the confinement factor $C_{\rm f}$ is unconstrained observationally.

\subsection{Implications for the Energy Budget}
\label{ssec:budget}

\begin{table*}
\begin{center}
	\caption{Inferred number densities, luminosities, and confinement factors}
	 \label{tab:Lum}
	\begin{tabular}{ lccccccccr}
	Name & $n_{\rm X}$ [$\rm{cm^{-3}}$] & $L_{\rm cool}/L_{\rm w}$ & $L_{\rm mech}/L_{\rm w}$ & $L_{\rm cond}/L_{\rm w}$ 
		& $L_{\rm dust}/L_{\rm w}$ & $C_{\rm f}\tablenotemark{a}$  & $C_{\rm f,\; all}$\tablenotemark{b} \\
	\hline
	\hline
	30 Doradus & 0.058 & 0.37\% & 15\% & $<97\%$ & $<40\%$ & $>0.84$ & --\\
	
	Carina & 0.14 &0.16\% & 4.3\% & $<22\%$ & $<11\%$ & $>0.36$ & $<0.58$\\
	
	NGC 3603 & 0.13 & 0.10\% & 3.7\% & $<41\%$ & $<11\%$  & $ >0.36$ & $<0.70$\\
	
	M17 & 0.27 & 0.55\% & 38\% & $<392\%$ & $<48\%$ & $>0.95$ & --  \\
	\hline
	\end{tabular}
\tablecomments{
For each \hii\ region in the sample, $n_{\rm X}$ is the number density inferred from the observed X-ray luminosity and best-fitting temperature. The columns $L_{\rm cool}/L_{\rm w}$, $L_{\rm mech}/L_{\rm w}$, $L_{\rm cond}/L_{\rm w}$, and $L_{\rm dust}/L_{\rm w}$ show the radiative cooling, mechanical work, conduction, and dust cooling luminosities normalized to the wind energy injection rate; the latter two are upper limits. Finally, $C_{\rm f}$ ($C_{\rm f, \; all}$) is the confinement factor that would be required to remove the unaccounted-for wind energy via bulk motion.\\
}

\tablenotetext{a}{Derived $C_{\rm f}$ includes energy loss due to mechanical work and radiative cooling. These act as lower limits since the values obtained for these energy loss mechanisms are reasonable estimates.}
\tablenotetext{b}{Derived $C_{\rm f, \; all}$ includes energy loss due to mechanical work, radiative cooling, thermal conduction, and dust heating via collisions. These act as upper limits since the values obtained for thermal conduction and dust heating via collisions are likely overestimates.}

\end{center}
\end{table*}

\begin{figure*}
\centerline{\includegraphics[trim=0cm 0cm 0cm 0cm,clip,width=0.6\textwidth]{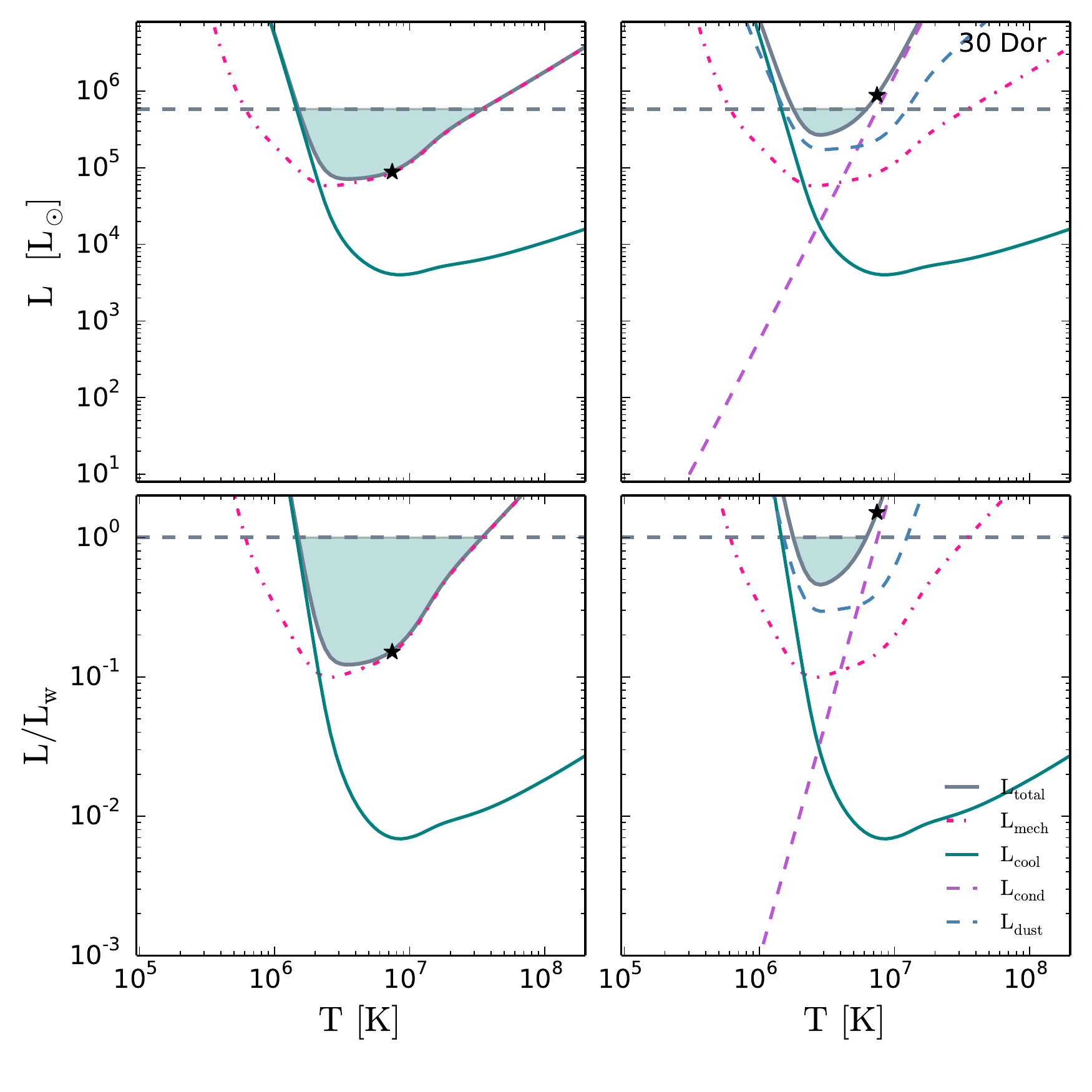}}%{30DorL11_LvsT4.pdf}}
\caption{
\label{fig:30Dor_LvsT}
Hot gas temperature versus the energy loss rates for the energy loss mechanisms described in Section 2 for the allowed hot gas number density for 30 Dor. The horizontal line in the top and bottom panels denote the stellar wind energy injection rate for 30 Dor. The left panels consider only $L_{\rm cool}$ and $L_{\rm mech}$, since these values are reasonable estimates whereas the left panels also include thermal conduction and dust heating via collisions, which are likely overestimates. Stars denote the values of $T_{\rm X}$ inferred from the shape of the X-ray spectrum.
}
\end{figure*}

\begin{figure*}
\centerline{\includegraphics[trim=0cm 0cm 0cm 0cm,clip,width=0.6\textwidth]{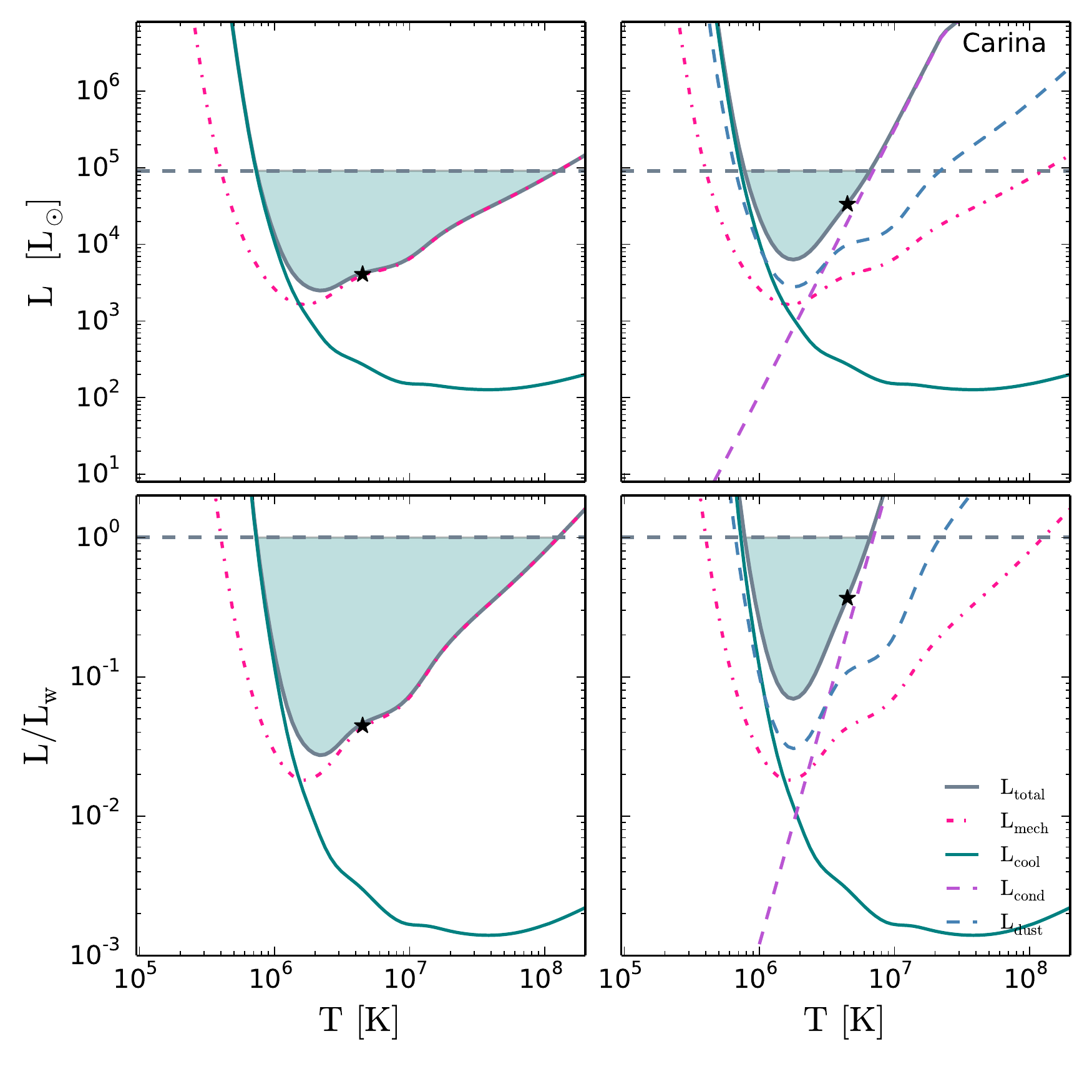}}%{CarinaT11_LvsT4.pdf}}
\caption{
\label{fig:Carina_LvsT}
Same as Figure \ref{fig:30Dor_LvsT} but for Carina.
}
\end{figure*}

In order to better constrain the dominant source of kinetic energy removal, and to illustrate the problem of the missing wind energy, we next calculate the various energy sinks as a function of the hot gas temperature. We perform this calculation at each temperature $T$ by using the observed X-ray luminosity to calculate the corresponding density $n_{\rm X,obs}$ from equation \ref{eq:nx}. For each $T - n_{\rm X,obs}$ pair, we then compute all the energy sinks discussed in the previous Section: radiative cooling, mechanical work, thermal conduction, and dust heating via collisions, and compare the sum of these cooling rates to the wind energy input rate.

\begin{figure*}
\centerline{\includegraphics[trim=0cm 0cm 0cm 0cm,clip,width=0.6\textwidth]{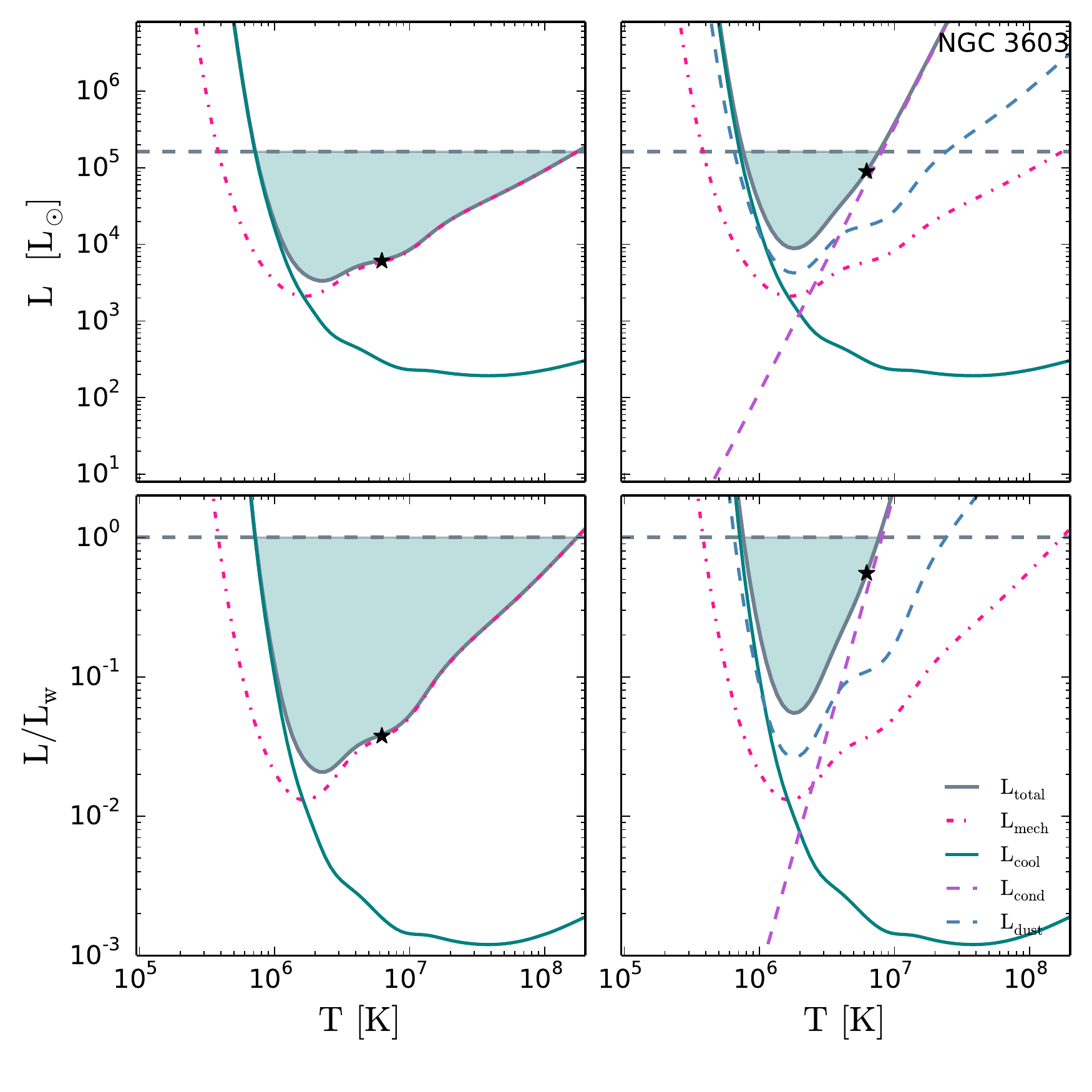}}%{NGC3603T11_LvsT4.pdf}}
\caption{
\label{fig:NGC3603_LvsT}
Same as Figure \ref{fig:30Dor_LvsT} but for NGC 3603.
}
\end{figure*}

\begin{figure*}
\centerline{\includegraphics[trim=0cm 0cm 0cm 0cm,clip,width=0.6\textwidth]{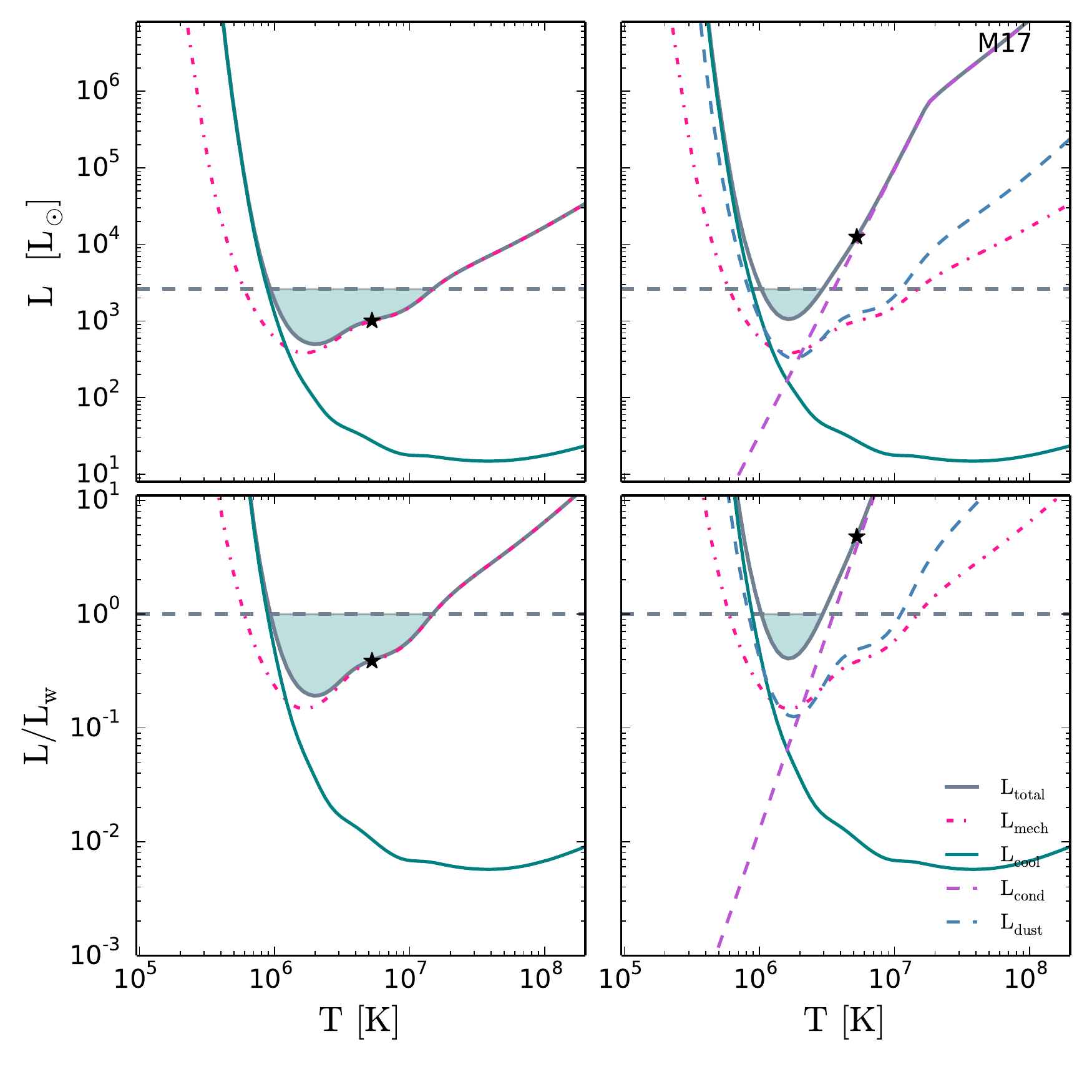}}%{M17T11_LvsT4.pdf}}
\caption{
\label{fig:M17_LvsT}
Same as Figure \ref{fig:30Dor_LvsT} but for M17.
}
\end{figure*}

Figures \ref{fig:30Dor_LvsT} - \ref{fig:M17_LvsT} show the results. Given the uncertainties in the true rates of conductive and dust heating, we perform this calculation both excluding them (left panels) and including them (right panels). The top panels show the absolute values of the individual and total energy loss rates, whereas the bottom panels show the energy loss rates as a fraction of the total energy injection rate by stellar winds. We remind the reader that values above the horizontal lines in these figures are not allowed due to energy conservation. The shaded regions illustrate how much energy is missing, i.e., what fraction of the injected wind energy cannot be accounted for by the sum of the various sinks we have been able to calculate. We also report these values, using the temperatures inferred from fitting the X-ray spectra, in Table \ref{tab:Lum}.

For temperatures reasonably consistent with the observationally-inferred values ($4.5\times10^6 \; \rm{K} \lesssim T_{\rm{X,\, obs}} \lesssim 7.5 \times 10^6 \; \rm{K}$, c.f.~Table \ref{tab:HIIregions}), we find that radiative cooling acts as a negligible energy sink, contributing to $<1\%$ of the fractional energy loss for all \hii\ regions. Mechanical work accounts for $3.7\%-38\%$ of the energy injected by winds, and for $<15\%$ in three of our four sample regions. We find that mechanical work can account for 38\% of the stellar wind energy injected in M17. This large fraction of energy transferred to mechanical work is likely due to M17's small size. The inferred number density from the X-ray emission is inversely related to the \hii\ region volume (i.e., $n_{\rm X} \propto V^{-1/2}$), thus a smaller volume for a given X-ray luminosity and plasma temperature would yield a larger inferred number density and hot gas pressure.

As illustrated in the right panels of Figures \ref{fig:30Dor_LvsT} - \ref{fig:M17_LvsT}, the situation is different if we include dust heating and/or thermal conduction. By combining these energy sinks with mechanical and radiative losses in Carina and NGC 3603, we can account for 37\% and 55\% of the injected energy, respectively. In the remaining two \hii\ regions, setting the conductive and dust cooling rates to their maximum would lead to a luminosity greater than that injected by winds. However, this assumption is only true if magnetic fields do not inhibit conductive losses in any way and the dust to gas ratio is the same in the hot gas as in the cold ISM. Neither of those assumptions are likely to be true, as we discuss further in Section \ref{sec:discussion}, even at the order of magnitude level. Only in M17, where we find $L_{\rm cond}/L_{\rm w} \approx 4.8$ (see Table \ref{tab:Lum}), is there a significant margin of error. In all the other regions, if the failure of these assumptions were to reduce the real conductive and dust luminosities by even a factor of a few compared to our upper limit, we would no longer be able to account for all the injected wind energy.

\subsection{Ways Out: Where's the Missing Energy?}

\begin{figure*}
\centerline{\includegraphics[trim=0cm 0cm 0cm 0cm,clip,width=0.6\textwidth]{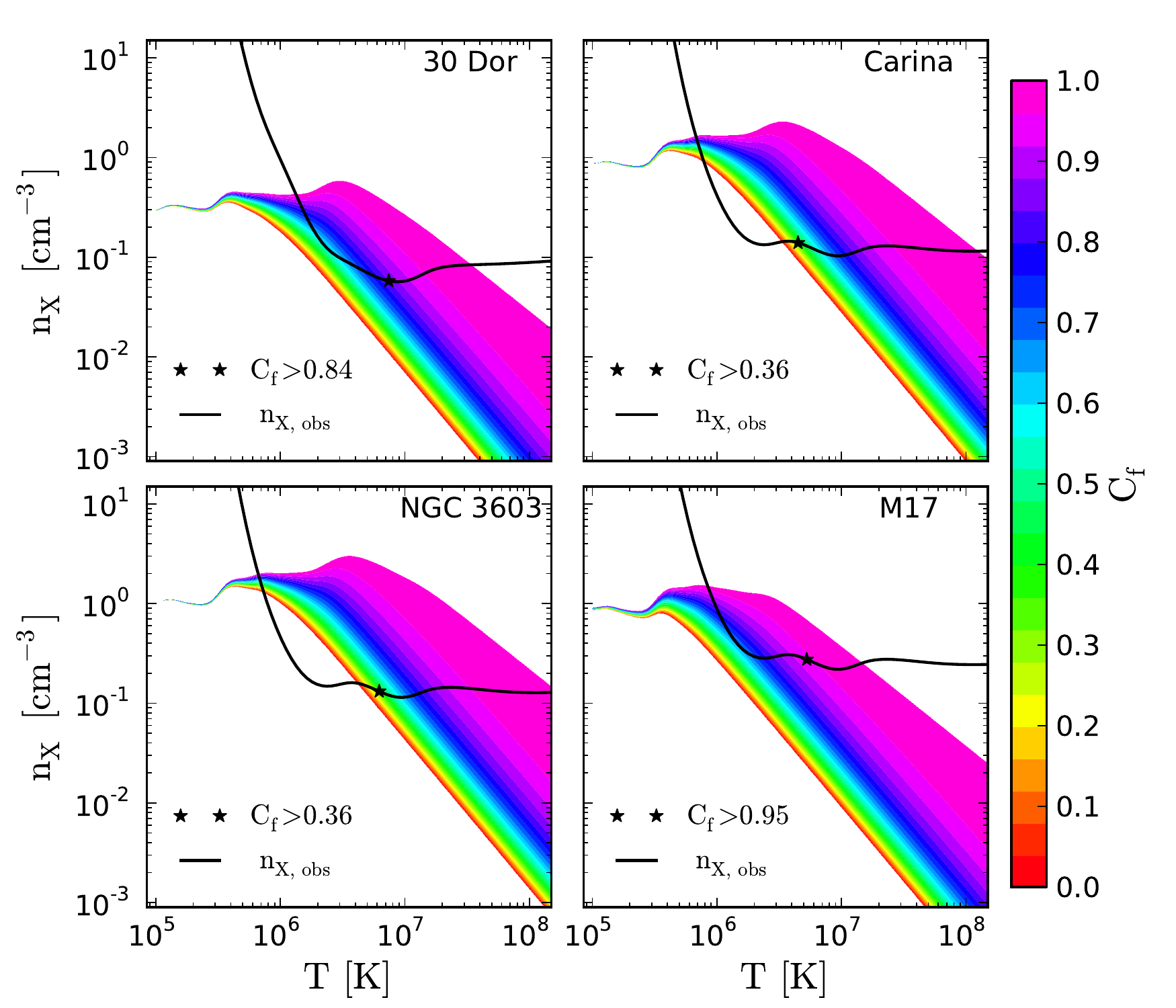}}%{AllCluster_Cf.pdf}}
\caption{
\label{fig:cf}
Contours of constant confinement parameter, $C_{\rm f}$, for all \hii\ regions in our sample. The value of $C_{\rm f}$ shown is that which would be required for physical leakage to account for all of the wind energy not removed by radiative cooling and mechanical work on the bubble shell. The curve of $n_{\rm X,obs}$ versus $T$ required for consistency with the observed X-ray emission is over-plotted. Stars denote the values of $T$ inferred from the X-ray spectra.
}
\end{figure*}

\begin{figure}
\centerline{\includegraphics[trim=0cm 0cm 0cm 0cm,clip,width=0.5\textwidth]{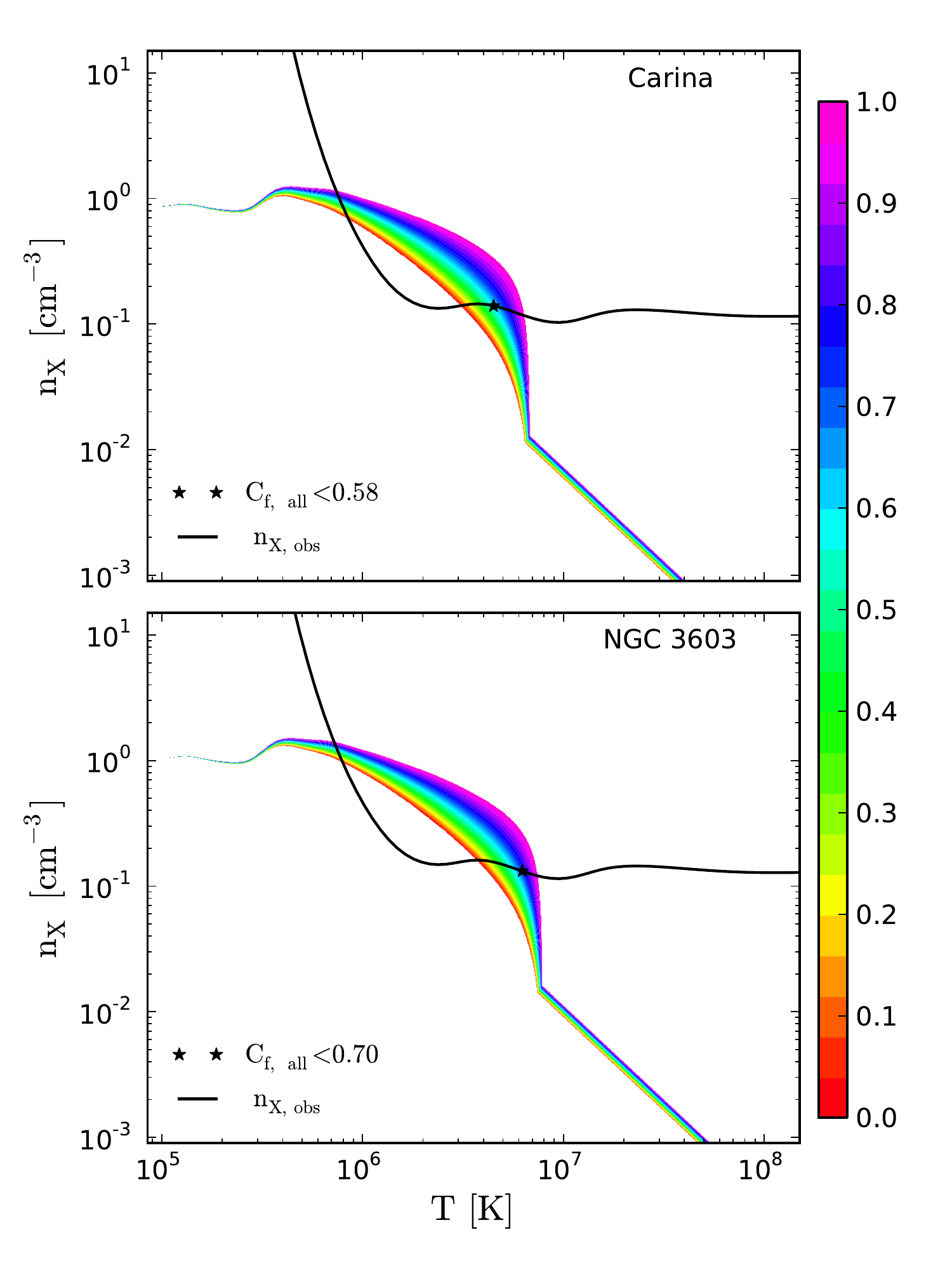}}%{Cf_allsink.pdf}}
\caption{
\label{fig:cf_allsink}
Same as Figure \ref{fig:cf} for Carina and NGC 3603 but also including the energy transfer associated with thermal conduction and collisional heating of dust grains.
}
\end{figure}

We have shown that the combined effects of radiative cooling and mechanical work cannot account for the missing energy of the hot post-shocked stellar wind material in our sample. Thermal conduction and dust heating via collisions might, but only if the assumptions described above are true. Are there other ways out?

One possible solution is physical leakage of the hot gas. If the bubble shell is porous, the hot gas can physically leak out since the sound speed of the hot gas is greater than the expansion rate of the bubble shell (as discussed in Section 2.5). The energy loss by physical leakage is controlled by the porosity of the bubble shell, which we can parameterize by the covering fraction $C_{\rm f}$. If the shell is very porous, the shock-heated gas will escape easily, resulting in a significant loss of the wind energy from the bubble, significantly reducing the X-ray luminosity. From the energy loss processes discussed in Section 2, we have that the total energy loss of the hot gas is
\begin{eqnarray}
\frac{dE}{dt} &=  L_{\rm{w}} - L_{\rm{cool}} - L_{\rm{mech}} - L_{\rm{cond}} - L_{\rm dust} - L_{\rm{leak}}.
\end{eqnarray}
\noindent
Using Equation~\ref{eq:Lleak} and assuming that these processes account for the total energy loss of the hot gas (i.e., $\frac{dE}{dt} = 0$), $C_{\rm{f}}$ is given by

\begin{equation}
\label{eq:Cf}
C_{\rm{f}} = 1 - \frac{2}{5}\frac{\left[  L_{\rm w} - L_{\rm cool} - L_{\rm mech} - L_{\rm cond} - L_{\rm dust}\right]}{4\pi R^2 \mu m_{\rm{p}} n_{\rm{X}} c_{\rm{s}}^3}
\end{equation}
\noindent
which depends on both $n_{\rm{X}}$ and $T$. 

Figure \ref{fig:cf} show contours of the values of $C_{\rm f}$ required to account for the missing energy as a function of hot gas density and temperature. To generate this figure, we solve equation \ref{eq:Cf} at each point in the $T - n_{\rm X}$ plane, assuming that only radiative cooling, mechanical work, and physical leakage contribute to energy loss, i.e., that $L_{\rm cond} = L_{\rm dust} = 0$. We also show the loci in the $T-n_{\rm X}$ inferred from the observed X-ray emission. The plot shows that physical leakage can adequately account for the missing energy for these \hii\ regions for plausible values of the confinement factor. For the observationally-favored temperature and the corresponding derived density values (denoted by points in the Figure), the required values of $C_{\rm f}$ are in the range $0.36-0.95$ (also see Table \ref{tab:Lum}). Adopting non-zero values of $L_{\rm cond}$ or $L_{\rm dust}$ would increase these values as can be seen in Figure \ref{fig:cf_allsink} and Table \ref{tab:Lum} for Carina and NGC 3603.

Finally, we note that there is one additional mechanism that we have not considered because we lack the ability to calculate it: turbulent mixing of the hot gas with cooler gas followed by conduction. As the shell expands into the ISM, the bubble shell interface becomes corrugated by instabilities \citep{strickland98}. These instabilities will lead to the addition of cooler, denser material in the bubble interior. This material can then mix with the hot gas, and the resulting large temperature variations over small scales will produce very rapid thermal conduction. For example, if the hot gas ($T\sim10^6-10^7$ K) mixes with the surrounding warm, ionized gas ($T\sim10^4$ K) the resulting mixture will have temperatures of $\sim10^5$ K  \citep{dunne03} and this gas will cool rapidly via metal-line cooling in the far-UV before adiabatically expanding and filling the whole \hii\ region. Figure \ref{fig:nvsT} shows that the cooling of the denser, mixed gas can effectively radiate all of the wind energy.

\section{Discussion}
\label{sec:discussion}

%In this section we discuss in more detail some of the subtle physical questions that enter our analysis of radiative, conductive, and dust cooling. %We defer the discussion of the implications of our results for the importance of stellar winds as a feedback mechanism regulating the formation of MSCs in the next Section.

\subsection{Deviations from Collisional Ionization Equilibrium}
\label{ssec:cie}

Throughout our analysis we have assumed that the post-shocked wind material responsible for the diffuse X-ray emission in \hii\ regions is in CIE. This assumption allows one to easily determine the allowed locus in the $T-n_{\rm X}$ plane for an optically thin plasma given its observed X-ray luminosity. This is because the ionization fractions of the plasma depend only on $T$ and $Z$ under the assumption of CIE, as does the emissivity, $j_{\rm \nu}$, and the radiative cooling function $\Lambda$. However, a hot plasma which was initially in equilibrium will deviate from equilibrium if the gas cools faster than it can recombine. The rapid cooling will cause the gas to become ``over-ionized" \citep{gnat07}. One such example is if the hot plasma undergoes rapid adiabatic expansion before significant radiative losses can occur \citep{breitschwerdt99} which has been observed in SN remnants \citep[][and references therein]{lalopez13b}. This scenario is likely the case for the hot gas we are considering in MSCs because the initial post-shock wind material will adiabatically expand and fill up the entire \hii\ region before suffering drastic radiative losses. 

\citet{gnat07} studied the time-dependent behavior of a hot, low-density plasma and found that non-equilibrium effects cause the radiative cooling rate to be suppressed by a factor of $2-4$ as compared to an equilibrium plasma. This result leads to an increase in the cooling time of a non-CIE plasma, rendering radiative cooling even more unimportant than our fiducial analysis suggests. Hence, our assumption of CIE likely only over-estimates the cooling rate for a given density.

Similarly, the emissivity, $j_{\rm \nu}$, integrated over the X-ray band, is of order $\Lambda$ for the temperatures that we consider (e.g., $T\gtrsim10^6$ K). Thus, if $\Lambda$ is suppressed by a factor of at most 4, then the integrated emissivity will decrease by a similar factor. This decrease in the emissivity will lead to the derived $n_{\rm X}$ to increase, at most, by a factor of 2, which would cause $L_{\rm mech}$ and $L_{\rm dust}$ to increase by factors of 2 and 4, respectively. Furthermore, $L_{\rm cool}$ will remain the same since the decrease in $\Lambda$ will cancel the increase in $n_{\rm X}$.  $L_{\rm cond}$ will remain approximately the same since the energy loss due to conduction in the unsaturated regime depends very weakly on density. Hence, we conclude that if the hot gas is not in CIE then radiative cooling is still an inefficient energy sink for the hot gas. The energy transfer due to mechanical work on the shell and dust heating via collisions will be larger than our fiducial estimates, but only by factors of order unity.

\subsection{Thermal Conduction and Magnetic Fields}
\label{ssec:magneticfields}

In our analysis, we found that thermal conduction can remove a significant amount of energy from the hot gas, but only if we assume that it is not substantially suppressed by the presence of a magnetic field oriented with field lines parallel to the wall of the bubble. If such a magnetic field is present, it inhibits electron transport between the hot and cold gas, reducing the conduction coefficient compared to our fiducial value by a factor of order $(r_{\rm e}/\ell_{\rm e})^2$, where $r_{\rm e}$ is the electron gyroradius and $\ell_{\rm e}$ is the electron mean free path. In a plasma of $10^7$ K gas with a density of $1$ cm$^{-3}$, roughly our observationally-inferred values, $\ell_e\sim 0.04$ pc. In comparison, for a magnetic field of strength $B$, the gyroradius is $r_e = 10^8 \sqrt{T_7/B_0}$ cm, where $T_7 = T/10^7$ K and $B_0 = B/1$ $\mu$G, so the ratio $(r_e/\ell_e)^2 \sim 10^{-20}$ even for an extremely weak field of $1$ $\mu$G. Thus even such a weak field will completely suppress conduction across field lines, and the only remaining question is the magnetic field geometry.

The molecular clouds out of which MSCs form are magnetized \citep{crutcher12a}. A number of authors  have simulated \hii\ regions expanding into magnetized media \citep[e.g.,][]{krumholz07b, wise11a, arthur11a, gendelev12a}, and a generic result of these simulations is that, as the \hii\ region expands, advection of material out of the low-density interior into the surrounding swept-up dense shell results in a decrease in field strength in the \hii\ region interior and an increase in field strength at the swept-up shell. This same phenomenon tends to reconfigure the field orientation such that, over most of the swept-up shell, the field is oriented parallel to the shell wall, i.e., the configuration that should be most effective at suppressing thermal conduction between the hot and cold phases. Although measuring magnetic fields in the shells that bound \hii\ regions is difficult, there is some observational evidence for this phenomenon operating. \citet{pellegrini07} measure the field strength in the photodissociation region (PDR) bounding the M17 \hii\ region to be $\sim 100$ $\mu$G, far above the mean interstellar value, suggesting that field amplification has taken place and that conduction is being suppressed. Indeed, \citet{dunne03} conclude that such a strong field is required to explain the observed low X-ray luminosity of M17.

Although M17 is just one example, and in it we have only circumstantial observational evidence that the field is oriented parallel to the dense shell wall, that plus the simulation results is highly suggestive that magnetic suppression of conduction is probably very effective in most H~\textsc{ii} regions, with the effectiveness depending on the detailed magnetic field configuration. It is therefore likely that conduction is not a significant source of energy loss. Note that this statement does \textit{not} apply to what we have termed turbulent mixing followed by conduction. The reason is that turbulence will cascade down to scales comparable to the electron gyroradius, and on such small scales conduction is no longer suppressed by magnetic fields. Magnetic fields suppress only laminar conduction, not conduction that is the end product of a turbulent cascade originating at an unstable interface.

\begin{figure}
\centerline{\includegraphics[trim=0cm 0cm 0cm 0cm,clip,width=0.45\textwidth]{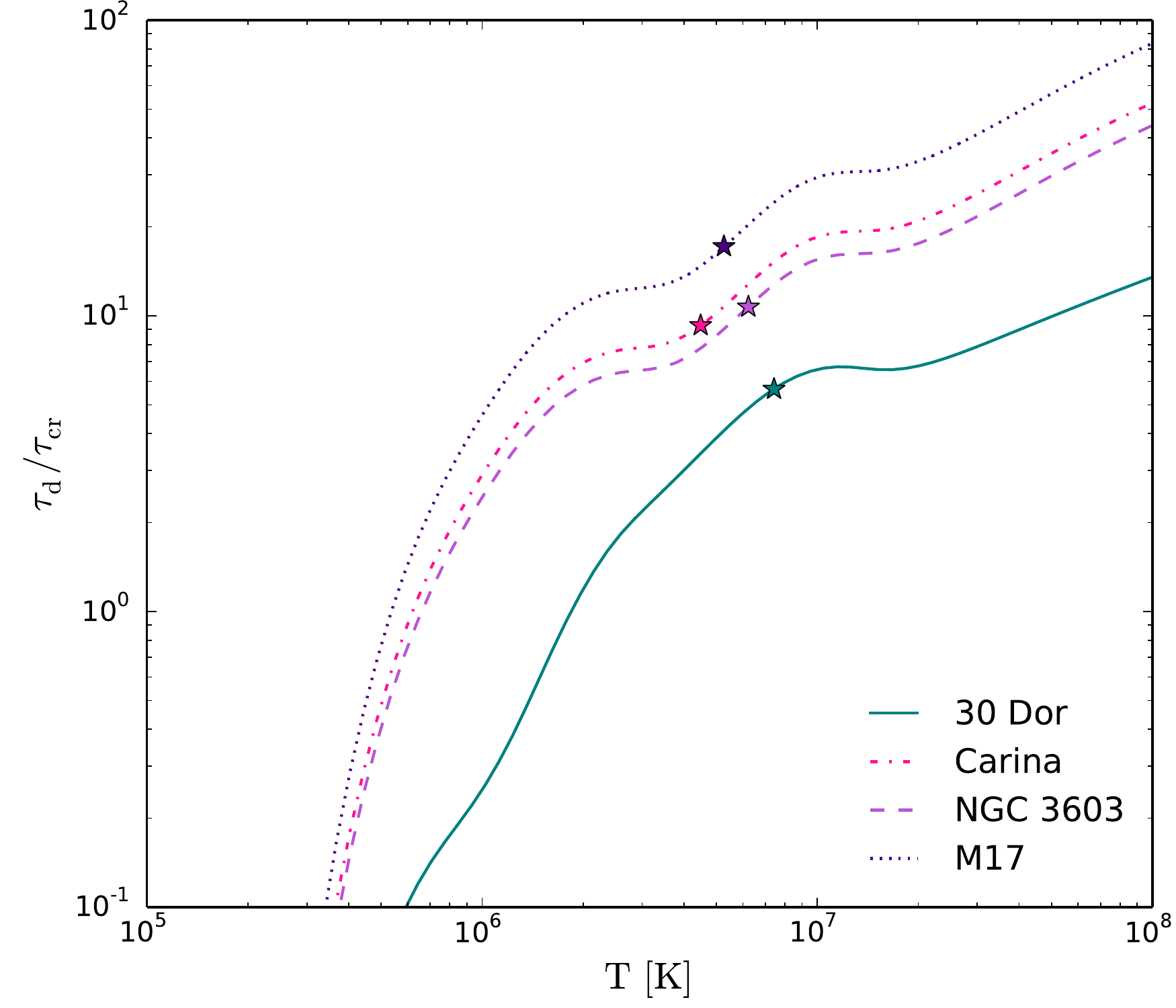}}%{dust_timeall.pdf}}
\caption{
\label{fig:dust}
Ratio of the dust destruction time scales to the crossing times for our \hii\ region sample. The points denote the observed $T_{\rm X}$ and the corresponding $n_{\rm X}$ values obtained from our derived $T-n_{\rm X}$ plane.
}
\end{figure}

\subsection{Dust Sputtering and the Dust Cooling Rate}
\label{ssec:dustlifetime}

We also found that the heating of dust by collisions with the hot electrons can be an important energy sink for the hot gas, but to be conservative we performed this calculation assuming the same dust to gas ratio in the hot gas as in the cold ISM. This assumption is unlikely to be satisfied. To see why, it is helpful to compare the mixing timescale of the dust that is entrained into the hot gas with the timescale for this dust to be destroyed by sputtering. The mixing timescale will be of order the crossing time, $\tau_{\rm cr} = R/c_{\rm s}$, of the hot gas. In comparison, the approximate lifetime of dust grains immersed in hot gas is
\begin{equation}
\label{eq:tdust}
\tau_{\rm d}  \approx 1\times 10^5 \left[ 1 + \left(\frac{T}{10^6 \; \rm{K}}\right)^{-3} \right]\left(\frac{a}{0.1\,\mu {\rm m}} \right) \left(\frac{n_{\rm i}}{{\rm cm}^{-3}}\right)^{-1} \rm{\; yr}
\end{equation}
where $a$ is the dust grain size \citep{draine11}.

Figure \ref{fig:dust} shows the ratio of dust grain lifetimes and crossing times for the \hii\ regions in our sample using our derived $n_{\rm X}$, where we have assumed a typical grain size of $0.1\;\mu$m in equation \ref{eq:tdust}. We find that under our derived conditions, the dust grains will survive from a few $\times 10^5$ years up to a couple Myrs. This results in the dust surviving from a few to $\sim 10$ crossing time scales for the temperatures given in Table \ref{tab:HIIregions}. This suggests that it is possible for dust grains with sizes greater than 0.1 $\mu$m to survive for some length of time in the hot gas. However, since the crossing and destruction timescales are not very different, our assumption that the dust abundance in the hot gas matches that in the cold gas is likely still a substantial overestimate. To keep the dust abundance so high, cold gas would have to be continually mixed into the hot \hii\ region interior on times not much greater than the hot gas crossing timescale. Such rapid mixing would likely in itself be a major cooling source, rendering the dust of secondary importance.

\section{Conclusions}
\label{sec:conclusion}

In this paper, we have examined the many different ways that MSCs can lose the kinetic energy injected by fast stellar winds from massive stars. These winds collide with each other and the ISM, generating hot shock-heated material at temperatures of $\sim10^7$ K, and the mechanical luminosity associated with the production of this gas is comparable to that provided by supernovae at later stages of stellar evolution. However, the effects of this gas on the ISM depend critically on where the energy ultimately ends up -- does it go into bulk motion of the cold ISM, possibly disrupting gas clouds and halting star formation? Is it radiated away as X-ray emission? Is it lost in some other way?

To address these questions, we have used the empirically determined properties from four LMC and MW MSCs. For each of these, the set of observational constraints is sufficient to allow us to estimate the wind energy input, and conversely, to estimate the rates of energy loss due to radiative cooling, mechanical work on the cool \hii\ region shell, thermal conduction, collisional dust heating, and physical leakage of hot gas out of the dense shell. We find that the radiative cooling of the hot gas accounts for less than 1\% of the total energy injected by stellar winds for the observed hot gas temperatures in the \hii\ regions we have considered. While this might appear to favor a significant fraction of the energy going into mechanical work and thus being available as a form of feedback, our estimates of the rate of mechanical work on the dense \hii\ region shell suggest that this is not the case. Instead, for all but one of the \hii\ regions (M17), we find that at most $\sim 15\%$ of the injected wind energy goes into doing mechanical work on the ISM. This limits the potential importance of winds as a stellar feedback mechanism, since it suggests that the efficiency with which they can be converted to bulk motion is fairly low.

This raises the question: if the bulk of the wind energy does not go into radiation nor mechanical work, where does it go? We identify four possible scenarios. The first is that the energy could be lost via thermal conduction at the hot-cold shell interface, followed by line radiation from this gas at far-UV wavelengths. However, this scenario appears to be viable only under the most optimistic possible assumptions. Thermal conduction will be dramatically reduced if there is a magnetic field parallel to the hot-cold interface, a configuration that simulations suggest should be common. It is possible to check this possibility via observations in several ways. If thermal conduction is the dominant loss mechanism, then observations of far-UV radiation in \hii\ regions should discover a significant mass of $\sim10^5$ K gas in these objects. More indirectly, polarization studies and Zeeman line splitting measurements of  the gas in \hii\ regions and their shells can allow one to determine the orientation of the magnetic field and magnetic field strength. If the magnetic field is indeed parallel to the hot-cold interface, then thermal conduction will be strongly suppressed. 

A second scenario is that wind energy stored in the hot gas is transferred to dust grains via collisions, and then radiated as infrared continuum. While this provides a sufficient energy sink to account for most of the injected wind energy if the dust content of the hot gas is the same as that of the cold gas, this too seems highly improbable. Grains $\sim 0.1$ $\mu$m in size will be destroyed by sputtering in the hot gas in a time that is only a factor of a few larger than the crossing timescale, which suggests that it would be difficult to maintain a large population of such grains. Observationally, one might be able to evaluate this possibility by checking for distortions in the dust continuum spectrum. Since sputtering will preferentially destroy small grains, the infrared spectral energy distribution (SED) produced by the remaining grains should be shifted to longer wavelengths than the usual dust SED.

The third way the energy can be accounted for is if the hot gas physically leaks out of the \hii\ region through holes in the bubble shell. These holes can be a result of stellar feedback punching holes in the dense \hii\ region shell or because the shell expands into a non-uniform ISM. We find that, for plausible values of the confinement factor of the dense shell, this loss mechanism would be sufficient to account for the missing energy. In support of this scenario, \citet{rp13} simulate the interaction of the mechanical energy input by stellar winds of three O-stars in a GMC and find that the hot gas generated by the shock heated stellar winds flows out of the GMC through low-density channels.

Our fourth and final scenario is that the hot gas can lose a significant amount of energy by mixing with the cold gas, followed by thermal conduction at the turbulent interface between the two -- turbulent conduction. The resulting mixed gas will have temperatures of $\sim10^5$ K and will drastically cool via radiation in the far-UV. There is one indirect piece of observational evidence for this scenario: \citet{bowen08} report high \ovi\ absorption in Carina, suggesting an overabundance of $\sim10^5$ K gas as compared to the normal ISM in the MW. Such an excess might also be evidence of laminar conduction without turbulent mixing, and one can distinguish between these scenarios by measuring the magnetic field strength and orientation. If a magnetic field parallel to the hot-cold interface is present, then the energy loss will most likely be dominated by turbulent conduction. We conclude that the either this scenario or physical leakage is the most likely explanation for the missing energy. %These results can have profound effects on the importance of stellar winds as a feedback mechanism in controlling \hii\ region dynamics.

These four possible scenarios suggest that one productive avenue for further investigation is three-dimensional simulations of stellar wind feedback. Simulations of wind feedback including self-gravity and a realistically-turbulent confining molecular medium are quite rare. \citet{rp13} is one of the few examples. However, even these simulations include none of the physical mechanisms -- magnetic fields, thermal conduction, dust sputtering -- that would be required to address any scenario except bulk leakage. Incorporating these mechanisms into future simulations would be a valuable complement to observational studies such as this one, and might lead to the development of new observational diagnostics that could be used to track down the missing energy.

\section*{Acknowledgements}
 ALR, MRK, and ERR acknowledge support from NASA through Hubble Archival Research grant HST-AR-13265.02-A issued by the Space Telescope Science Institute, which is operated by the Association of Universities for Research in Astronomy, Inc., under NASA contract NAS 5-26555. Support for this work was also provided by NASA through Chandra Award Number GO213003A and through Smithsonian Astrophysical Observatory contract SV373016 to MIT and UCSC issued by the Chandra X-ray Observatory Center, which is operated by the Smithsonian Astrophysical Observatory for and on behalf of NASA under contract NAS803060. ALR acknowledges support from the NSF Graduate Research Fellowship Program. MRK acknowledges support from NSF grant AST-0955300, and from NASA ATP grant NNX13AB84G. ERR acknowledges support from the David and Lucile Packard Foundation and NSF grant: AST-0847563.  Support for LAL was provided by NASA through the Einstein Fellowship Program, grant PF1120085, and the MIT Pappalardo Fellowship in Physics. ALR thanks Chris McKee, Jason X. Prochaska, Leisa Townsley, and Jorick Vink for useful discussions.
\bibliographystyle{mn2e_fix}
\bibliography{refs}
\end{document}